# Ultrashort spin-orbit torque generated by femtosecond laser pulses

T. Janda,[1,2,a] T. Ostatnický,[1] P. Němec,[1] E. Schmoranzerová,[1] R. Campion,[3] V. Hills,[3] V. Novák,[4] Z. Šobáň,[4] and J. Wunderlich[2,4]

[1]*Faculty of Mathematics and Physics, Charles University, Ke Karlovu 3, 121 16 Prague 2, Czech Republic*

[2]*Institute for Experimental and Applied Physics, University of Regensburg, Universitätsstr. 31, 93053 Regensburg, Germany*

[3]*School of Physics and Astronomy, University of Nottingham, Nottingham, NG7 2RD, UK*

[4]*Institute of Physics ASCR, v.v.i., Cukrovarnická 10, 162 00 Prague 6, Czech Republic*

**ABSTRACT**

**To realize the very objective of spintronics, namely the development of ultra-high frequency and energy-efficient electronic devices, an ultrafast and scalable approach to switch magnetic bits is required. Magnetization switching with spin currents generated by the spin-orbit interaction at ferromagnetic/non-magnetic interfaces is one of such scalable approaches, where the ultimate switching speed is limited by the Larmor precession frequency. Understanding the magnetization precession dynamics induced by spin-orbit torques (SOTs) is therefore of great importance. Here we demonstrate generation of ultrashort SOT pulses that excite Larmor precession at an epitaxial Fe/GaAs interface by converting femtosecond laser pulses into high-amplitude current pulses in an electrically biased *p-i-n* photodiode. We control the polarity, amplitude, and duration of the current pulses and, most importantly, also their propagation direction with respect to the crystal orientation. The SOT origin of the excited Larmor precession was revealed by a detailed analysis of the precession phase and amplitude at different experimental conditions.**

## INTRODUCTION

The ever-increasing demand for a faster and low-energy-consumption electronics calls for a development of electronic devices with high operational frequencies. Spintronics is among the most frequently considered solutions, and currently represents a highly active and rapidly developing field placed at the intersection of relativistic quantum physics, materials science, and nanoelectronics. Commercial spin-based devices such as hard disk drives and magnetic random-access memories are intrinsically robust against charge perturbations and non-volatile. These applications rely on two opposite magnetisation orientations representing logic "zero" and "one" and the fastest speed of switching magnetization between the two orientations is limited by the Larmor precession. Magnetization reversal has been realized by transferring linear momentum into spin angular momentum through the spin Hall effect (SHE) and the inverse spin galvanic effect (iSGE), and the underlying spin-orbit fields have been studied extensively in ferromagnetic (FM) / non-magnetic (NM) systems. In these bilayer systems, a spin current generated in the NM bulk or at the FM / NM interface is absorbed in the adjacent FM layer, and the corresponding spin-orbit fields are, therefore, primarily interfacial effects [1, 2, 3, 4, 5]. Epitaxial interfaces are of particular interest, since effective spin-orbit fields may depend on the current direction with respect to crystallographic orientation [6, 7, 8].

---

[a] Electronic mail: tomas.janda@ur.de

In this work, we demonstrate the excitation of Larmor precession at an epitaxial iron-gallium arsenide (Fe/GaAs) interface by ultrashort (sub-picosecond) spin-orbit torque (SOT) pulses. We use a magneto-photo-electric device, where the Fe is epitaxially grown on a silicon-doped n-GaAs / intrinsic GaAs / carbon-doped p-GaAs photodiode structure. The metal/semiconductor interface is forming a Schottky barrier in series with the *p-i-n* photodiode. Irradiation of the electrically biased Fe/*n-i-p*-GaAs heterostructure with femtosecond laser pulses can generate ultrashort sub-ps current pulses with up to ~ 100 mA large amplitudes that propagate along the GaAs/Fe interface. The magnitude and polarity of the applied bias voltage allow to control the amplitude and polarity as well as the duration of the current pulses. In addition, as it is particularly important for this study, we can also control the lateral propagation direction of the current pulses with respect to both the GaAs crystal orientation and the magnetic easy axes of the Fe film. We realize this by focusing the excitation laser pulse to a specific position in our disk-shaped device. By carefully analyzing phase and amplitude of the induced Larmor precession we identify conditions where its excitation is dominated by SOT pulses.

**RESULTS**

**A. Studied sample**

In Fig. 1(a) we show a sketch of our GaAs *p-i-n*-diode grown on a semi-insulating GaAs substrate. On top of the diode (on the *n*-doped layer) an ultrathin (2 nm ≈ 14 monolayers) ferromagnetic Fe film is deposited and protected by an aluminum-oxide layer (see Supplementary Fig. S.3 and Note 3 for details on magnetic properties of the studied sample). The GaAs diode consists of a 670 nm thick *p*-doped layer (carbon-doped, $n_C = 2\times10^{18}$ cm$^{-3}$), a 1000 nm thick intrinsic layer, and three *n*-doped layers with a silicon dopant concentration gradually increasing towards the diode surface: 150 nm thick layer with $n_{Si} = 1\times10^{17}$ cm$^{-3}$, 15 nm thick layer with $n_{Si}$ ranging from $1\times10^{17}$ to $5\times10^{18}$ cm$^{-3}$, and 15 nm thick layer with $n_{Si} = 5\times10^{18}$ cm$^{-3}$. The *p-i-n* / Fe stack was patterned into a disc with a diameter of 100 μm, as shown in the micrograph in Fig. 1(b). Here, the circularly shaped 100 nm thick Au-contact (yellow) is used to electrically connect the thin Fe film (light grey), which is separated from the rest of the layer by a trench (dark grey). The built-in electric field *E*, which is present in the intrinsic layer of the diode, is superimposed by an additional field generated by an applied bias voltage between the two Au contacts connecting both the *p*-doped GaAs and the Fe film on top. Note that the forward (reverse) voltage is positive (negative) in our convention. As shown in Supplementary Fig. S.4(a), without laser irradiation the diode can withstand a large reverse bias voltage with negligible leakage current. Moreover, we also can apply a relatively large forward voltage of up to +8 V without incurring large dark currents of more than 200 μA due to the formation of an additional rectifying Schottky barrier at the Fe/*n*-GaAs interface [9]. In summary, the currents without laser irradiation are at least 3 orders of magnitude smaller than the amplitudes of the current pulses excited by fs laser pulses over the entire bias voltage range applied in our experiments.

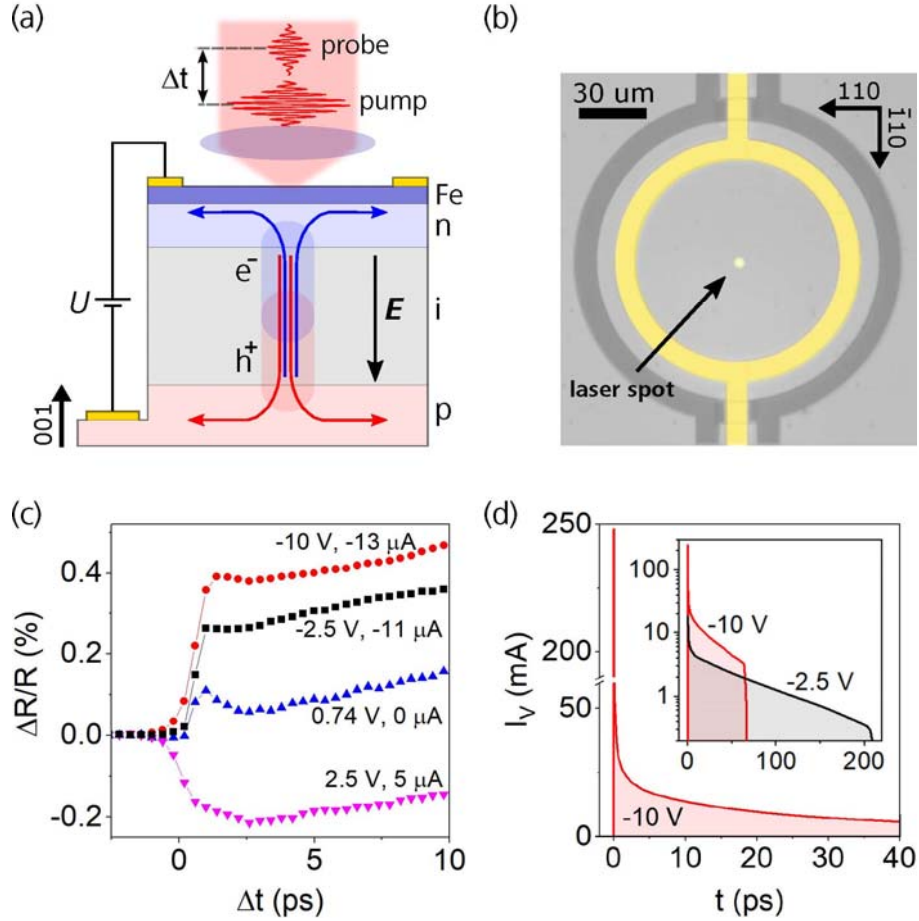

**Fig. 1 Device structure and properties of the photogenerated current pulses. (a)** Schematic illustration of the GaAs *p-i-n*-diode structure with a thin ferromagnetic Fe film deposited directly on top of the diode. After an impact of a femtosecond laser pulse, a cloud of electron-hole pairs is generated. The electrons ($e^-$) and holes ($h^+$) are separated by a strong electric field $E$ in the intrinsic layer (*i*) and accelerated towards the oppositely-charged electrodes creating a short current pulse, whose direction is schematically shown by arrows. To measure the dynamics of excited charge, spin and magnetization, a second time-delayed laser pulse can be used. **(b)** Micrograph of the photodiode lateral structure. The studied Fe/GaAs diode (light grey) is isolated from the rest of the sample by a trench (dark grey). The Fe film is electrically contacted by Au circular contact (yellow), which bridges the trench. The Au contact to the bottom *p*-GaAs electrode is outside of the displayed image. The bright spot in the image center is the focused laser spot. **(c)** Transient reflectivity measured in a pump-probe experiment for different applied bias voltages indicating the sub-picosecond photocurrent onset. The measured photocurrent averaged over the period of the stroboscopic experiment is shown for each bias. **(d)** Numerical simulation of the vertically propagating photocurrent pulse entering the *n*-GaAs/Fe electrode for an external bias of -10 V. The initial current spike decays at a timescale comparable to the laser pulse duration of ≈ 100 fs. Inset: The subsequent charge dynamics is much slower due to the electrical screening with exponential decay times of ≈ 40 ps (80 ps) for -10 V (-2.5 V); note that logarithmic *y*-scale is used in the Inset.

## B. Temporal characteristics of the photocurrent pulse

In order to produce ultrashort electrical current pulses of large amplitudes, the electrically biased *p-i-n* / Fe structure is illuminated by a focused femtosecond laser beam. Absorption of the laser pulse in the μm-thick intrinsic layer creates electron-hole pairs which are instantaneously separated by the large internal electric field and accelerated towards the opposite electrodes creating an ultrashort current pulse, as schematically shown in Fig. 1(a). Two stroboscopic time-resolved techniques were applied to explore the temporal characteristics of the current pulses generated in our photodiode. Pump-probe reflectivity measurements where used to identify the sub-picosecond onset of the current pulses. Photocurrent correlation measurements were used to monitor the time characteristics of the decaying tail of the pulses on the scale of tens of picoseconds. These two techniques, therefore, provide complementary information about the electrical pulses. For more information see Methods section.

Fig. 1(c) shows differential reflectivity data measured in the pump-probe experiment, i.e., evolution of the sample reflectivity triggered by an impact of the pump laser pulse at time delay $\Delta t = 0$. The pump and probe laser spots with a diameter of $\approx 1$ μm were overlapped in the center of the device. Differential reflectivity, which is connected with the transient photo-generated charge carriers, has been previously used to estimate the electrical pulse duration [9]. Here it provides information about the onset of the electrical pulse. To interpret the measured differential reflectivity data, we first consider the case where the built-in electric field of the *p-i-n* diode is almost compensated by an applied voltage bias causing *zero* averaged photocurrent. In this case, the transient reflectivity varies on the fast-scale only by a small amount. With a *non-zero* averaged photocurrent and a corresponding non-zero internal electric field, the onset of the transient reflectivity variation increases with increasing magnitude of the applied bias and the sign of the variation switches when the polarity of the photocurrent changes [see Fig. 1(c)]. Most importantly, the reflectivity changes within a sub-picosecond range. We attribute the bias-dependent reflectivity variation to the Franz-Keldysh effect [10, 11], as the photoexcited charge carriers separate from each other very quickly and screen the internal electric field accordingly. The fast rising and falling current until the screening has been established corresponds to an ultrafast initial current spike. Its duration is, therefore, equal to the time span of the measured reflectivity variation.

The sub-picosecond time scale of the current spikes identified by the transient reflectivity measurements is also confirmed by theoretical modeling, see Fig. 1(d). The simulation shows that the photoexcited electrons and holes in the intrinsic region of the diode are accelerated by the strong internal electric field towards the opposite electrodes thus causing an immediate onset of the photocurrent (see Supplementary Note 6 for details). As the two oppositely charged carrier clouds start to separate, the associated electric field pointing against the externally applied bias arises at a sub-picosecond timescale causing the rapid sub-picosecond decay of the current. The corresponding current spikes reach amplitudes of hundreds of mA. As apparent in Fig. 1(d), following the initial sub-ps current spike there is a further slower decay of the flowing current, which proceeds at a timescale of tens of picoseconds. It results from a drainage of the remaining photo-charges and a restoration of the internal strong electric field and it corresponds to the current pulse tail. In Fig. 1(d) we show the simulated current pulse flowing vertically (i.e., perpendicularly to the sample plane) in the micrometer-wide laser-pulse-illuminated channel at -10 V. As shown in the inset, the current pulse tail becomes longer for smaller biases and it is terminated when all the photogenerated charge is drained away from the illuminated spot. A detailed description of the performed numerical simulations can be found in Supplementary Notes 6 and 7.

The decay of the current pulse tails in the presence of the screening electric field was experimentally studied by the photocurrent correlation measurements for various applied biases (see Supplementary Fig. S.5 and Note 5). For $\approx 1$ μm laser spot size, decay times ranging from 20 ps at -10 V applied bias to $\approx 175$ ps at zero applied bias, where the photocarriers are accelerated only by the built-in field at the *p-i-n* junction, were measured [see Supplementary Fig. S.5(a) and Fig. 4(d)]. For larger laser spots, the decay times become longer due to the presence of electric screening in the wider photoexcited area [see Supplementary Fig. S.5(b)]. Comparing the Larmor frequencies of the thin Fe layer, which are tens of GHz [see Fig. S.3(b)], with the temporal characteristics of the current pulses generated by femtosecond laser pulses in our device, we conclude that the observed Larmor precession is excited mainly during the 10-picosecond tail of the current pulse. Nevertheless, the sub-picosecond current spike, with its high current amplitude, could become relevant for excitation of antiferromagnetically ordered materials where the precession frequencies reach the THz scale [12, 13, 14].

### C. Magnetization dynamics induced by spin-orbit torque pulses

To study the magnetization dynamics induced by spin-orbit torque pulses, we need to consider not only the temporal properties of the generated current pulse, but also its propagation direction. Due to our vertical photodiode design, the photocurrent pulses can be generated locally at any position within the disk-shaped pillar structure by positioning the focused laser spot inside the annular Au contact. As illustrated in Fig. 2(b) for 3 different spot positions, the net flux of the laterally propagating current pulse is always directed towards the shortest distance to the annular Au contact providing a unique and simple opportunity to control the orientation of the lateral current pulses in only one single device.

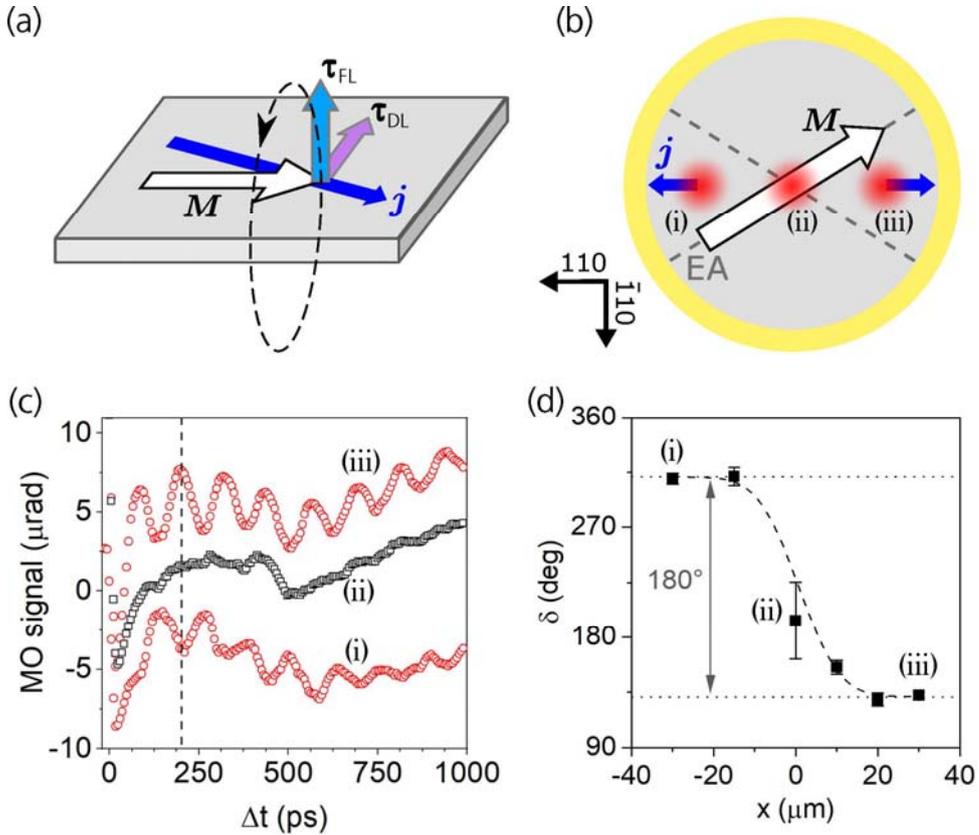

**Fig. 2 Magnetization precession induced by ultrashort SOT pulse. (a)** Schematic illustration of the two possible SOT contributions acting on magnetization ($M$) due to the photocurrent pulse ($j$): field-like ($\tau_{FL}$) and damping-like ($\tau_{DL}$) torques. The two orthogonal torques lead to magnetization precessions (dashed ellipse) that differ by 90° in the initial phase. **(b)** Sketch of the photodiode with three positions located in the device center and ±30 μm away from the center along the [110] crystallographic direction, which were investigated by the spatially-overlapped pump and probe laser spots. The net photocurrents in positions (i) and (iii) are opposite due to the location of the annular Au contact. No net lateral photocurrent is generated in position (ii). The experiment was performed with no external magnetic field applied when $M$ was oriented along one of the easy axes (EA), which are indicated by the dashed lines. **(c)** Magnetization precession measured in the three positions indicated in (b) for applied bias of -10 V; the curves are vertically shifted for clarity. The precessional signal is correlated with the polarity and magnitude of the lateral current pulse. **(d)** Points: initial precessional phase $\delta$ as a function of the laser spots position $x$ within the photodiode. The value $\delta \approx 135°$ for $x > 0$ indicates the presence of damping-like SOT, as discussed in the text. Dashed line: fit by a step-like function smeared out by the Gaussian laser beam profile with $\approx 25$ μm diameter.

As described in detail in Supplementary Note 6, the photocurrent pulse generated in the device initially flows in the vertical direction and after reaching the top Fe/*n*-GaAs electrode it propagates laterally toward the annular Au contact. To eliminate the effect of the Oersted magnetic field generated by the vertically propagating current pulse, which was investigated in Ref. 9, we performed experiments with pump and probe laser spots focused at the same position. To achieve a precise spatial overlap of pump and probe beams, we used relatively wide laser spots with a diameter of about 25 µm (see Methods section for more details). The radially symmetric distribution of the Oersted field generated by the vertical current in the pump-illuminated area averages out to zero within the spatially overlapped probing area and, therefore, a possible magnetization precession arises only from torques generated by the lateral current.

The lateral current pulse can act on the iron magnetization again via the associated Oersted magnetic field, which is a *non-local* effect. Therefore, the field generated by the electron current flowing in the top n-doped GaAs layer is compensated to a large extent by the hole current flowing in the bottom p-doped GaAs layer as both types of charge carriers move in the same direction coupled by the Coulomb interaction (see Supplementary Note 6). On the other hand, the spin-orbit torque is a *local* effect originating from the Fe/GaAs interface and, therefore, it is generated only by the electron current. In summary, a possible magnetization precession observed in the configuration with spatially overlapped pump and probe laser spots can only be triggered by SOT generated by the lateral photocurrent pulse. Other current pulse-unrelated effects that could in principle excite magnetization precession can be excluded in our experiments, as discussed in detail in Supplementary Note 1.

The spin-orbit field arises from a broken inversion symmetry both due to the epitaxial Fe/GaAs interface and due to the missing inversion centre of the zinc-blende crystal structure of the GaAs bulk [15, 16]. The respective contributions, the Bychkov-Rashba-like field $\boldsymbol{H}_R \sim (k_x \boldsymbol{e}_y - k_y \boldsymbol{e}_x)$ and the Dresselhaus-like field $\boldsymbol{H}_D \sim (k_y \boldsymbol{e}_y - k_x \boldsymbol{e}_x)$, lie in the plane of the Fe/GaAs interface and they depend linearly on the lateral components of the electron linear momentum $\hbar\boldsymbol{k}$ [8]; see also Fig. 3(b). Here, $\boldsymbol{e}_x$ and $\boldsymbol{e}_y$ are unit vectors pointing along [100] and [010] crystallographic directions of our epitaxially grown Fe/GaAs structure, respectively. Hence, when a lateral charge current is applied, Bychkov-Rashba- and Dresselhaus-like fields superimpose to an effective spin-orbit field which scales linearly with the current magnitude and whose magnitude and orientation also depend on the current direction; see Fig. 3(a). The current-induced spin-orbit field can cause two types of spin-orbit torques that act on the magnetization in two orthogonal directions: the field-like SOT ($\tau_{FL}$) and the damping-like SOT ($\tau_{DL}$) tilting the magnetization perpendicular and parallel to the sample plane, respectively, as depicted in Fig. 2(a). Accordingly, current pulse-induced magnetization precessions caused by the damping-like SOT only and by the field-like SOT only would be mutually phase shifted by 90°.

The direction of the lateral current pulse flowing in our photodiode from the pump laser spot towards the annular Au contact depends on the laser spot position, as depicted in Fig. 2(b). When the laser spot is focused in the center of the disc, the angular distribution of the lateral current flowing towards the ring-shaped contact is equal in all directions. Therefore, within the spatially overlapped probe laser spot, the net lateral current equals to zero and, consequently, no net spin-orbit torque is expected. In contrast, when the overlapped laser spots are moved out of the device center, the lateral current distribution becomes asymmetric and the current density is the largest in the direction of the spot-displacement since here the distance to the annular Au contact is the shortest. The corresponding net current flow directions are indicated in Fig. 2(b) by blue arrows.

In Fig. 2(c) we show magneto-optical (MO) pump-probe traces measured at three different positions indicated in Fig. 2(b). The measured dynamic MO signal is due to polar MO Kerr effect which is proportional to the perpendicular-to-plane magnetization component (see Supplementary Note 1 for details). While the precessional amplitude is almost zero at the disc center, Larmor precession of magnetization is clearly apparent when the spatially-overlapped pump and probe laser spots are displaced by ±30 μm away from the disc center along the [110] and [$\bar{1}$10] directions. Importantly, these precessions are mutually phase-shifted by 180°, as indicated by a vertical dashed line in Fig. 2(c). The presence (absence) of the magnetization precession and its opposite phases are clearly correlated with the presence (absence) and opposite directions of the lateral current pulses at the 3 different positions. This observation confirms that the precession is triggered by the lateral current pulses and it excludes any thermally-induced effects, such as temperature-induced magnetic anisotropy variations (see Supplementary Note 1 for a detailed discussion).

The measured pump-probe traces were fitted by a damped harmonic function

$$MO(\Delta t) = A\cos(2\pi f \Delta t + \delta)e^{-\frac{\Delta t}{\tau_d}} + P_4, \qquad (1)$$

where $A$, $f$, $\delta$, and $\tau_d$ are precessional amplitude, frequency, initial phase and damping time, respectively. $P_4$ is a 4th-order polynomial used to remove the signal background which is not related to the magnetization precession. In Fig. 2(d) we show the initial phase of the magnetization precession measured at different positions along the [110] crystallographic direction. As already mentioned, the phase changes by 180° when crossing the device center, i.e., when the current pulse propagation direction is inverted. The width of the transition region corresponds to the used laser beam size, which is ≈ 25 μm in this particular experiment. Importantly, the experimentally measured value of the initial phase $\delta$ is ≈ 135° (and 315°) for a given net-current direction. As we discuss in Supplementary Note 1, the initial phase of the magnetization precession depends not only on the direction of the corresponding torque, but also on the duration of the stimulus. In the measurements shown in Fig. 2, where the current pulse duration [see Supplementary Fig. S.5(b)] is significantly longer than the Larmor precession period of ≈ 100 ps [see Supplementary Fig. S.3(b)], a phase of ≈ 90° is expected for a precession excited by the current pulse-induced Oersted field. This effect, therefore, can't explain the observed phase of ≈ 135°, in agreement with the aforementioned strong suppression of the Oersted field effects in our experimental geometry. On the other hand, the measured precession phase can be explained by a superposition of the field-like and damping-like SOTs which would lead to precession phases of ≈ 90° and ≈ 180°, respectively.

To further investigate the importance of SOT relative to the Oersted-field torque for inducing the magnetization precession, we performed additional MO pump-probe experiments in a configuration with pump and probe laser spots being spatially separated. In this case, both SOT from the lateral current pulse and the Oersted-field torque from the vertical current pulse are present. The net direction of the lateral current pulse propagation within the probed area is determined by the position of the probe laser spot relative to the pump spot, which is the origin of the laterally propagating current pulse. To achieve the required spatial separation between the pump and probe laser spots within the device, we used rather small laser spot size of only ≈ 1 μm for this set of experiments. The tightly focused pump laser pulse excitation produced much shorter current pulses than in the previously described experiments with ≈ 25 μm wide laser spots, which is because of a reduced effect of charge-screening in the case of smaller spots [cf. Supplementary Figs. S.5(a) and S.5(b)]. We first investigated the dependence of the induced Larmor precession on the current pulse propagation direction relative to the Fe/GaAs crystallographic directions. The results of these experiments are summarized in Fig. 3. The sketch in Fig. 3(a) indicates the positions of the spatially-separated pump and probe laser spots, which are depicted by large and small circles, respectively. Each color represents a particular pump-probe experiment at the respective position on the disk-shaped device. The effective spin-orbit field $H_{SO}$, which is composed of the Bychkov-Rashba-like ($H_R$) and the Dresselhaus-like ($H_D$) fields [8], is schematically shown in Fig. 3(b) for different lateral photocurrent pulse directions. When the current pulse propagates along the [110] crystallographic direction, which corresponds to the situation when the probe laser spot is spatially displaced relative to the pump spot along this direction, $H_R$ and $H_D$ point along the same direction and, therefore, they sum up. On the contrary, for a current pulse propagating along the [$\bar{1}$10] direction, i.e., when the probe laser spot is displaced accordingly, $H_R$ and $H_D$ subtract, which then leads to a correspondingly smaller effective spin-orbit field $H_{SO}$, as indicated in Fig. 3(a) by a size of the black arrow.

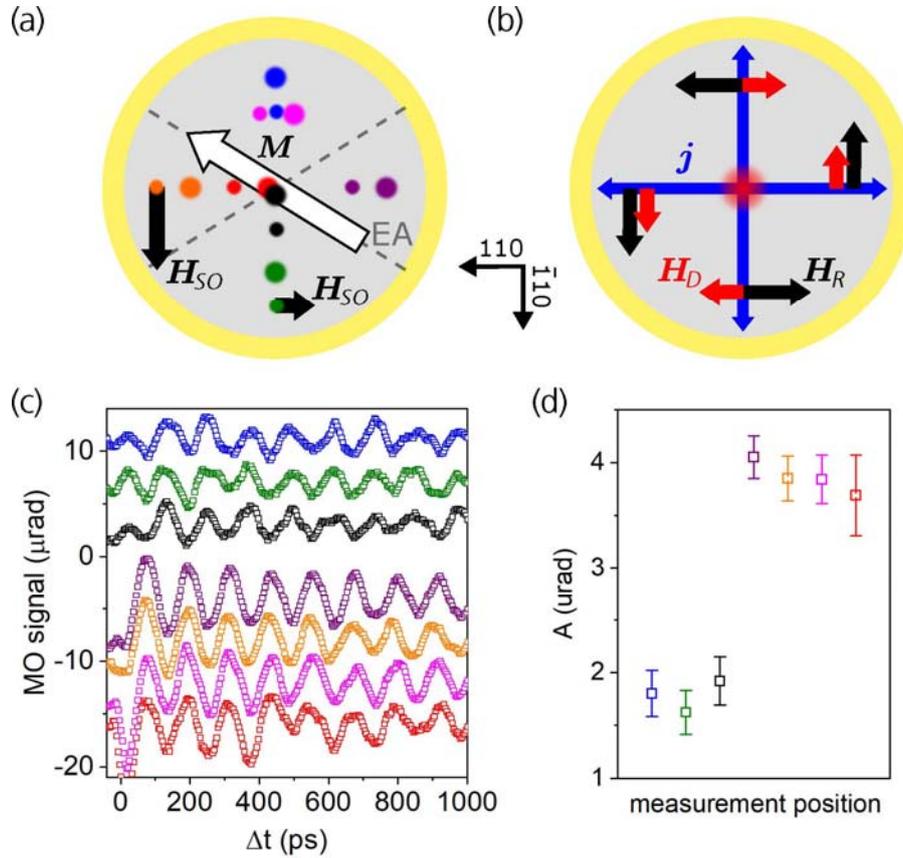

**Fig. 3 Anisotropy of the SOT-induced precession amplitude with respect to the photocurrent direction.** (a) Schematic illustration of the photodiode with positions of spatially-separated pump and probe laser spots indicated by different colors. The larger and smaller dots indicate the pump and probe spot positions, respectively (with a diameter of ≈ 1 μm), which were always 9 μm apart with the probe spot displaced either along [$\bar{1}$10] or [110] crystallographic direction. The directions of the magnetization $M$ (white arrow) and the local total spin-orbit field $H_{SO}$ (black arrows) are also indicated. (b) Directions of the Bychkov-Rashba ($H_R$, black arrows) and Dresselhaus ($H_D$, red arrows) contributions to the total spin-orbit field $H_{SO}$ for different photocurrent directions (blue arrows) in the device. Combination of the two anisotropic contributions results in different magnitudes of $H_{SO}$ when probing along [$\bar{1}$10] and [110] in (a). (c) Magnetization precession measured for applied bias of -10 V at positions indicated in (a) by the corresponding colors; the curves are vertically shifted for clarity. (d) Precessional amplitude $A$ extracted by fitting the data in part (c) by Eq. (1) for $\Delta t > 100$ ps. The colors of the data points correspond to that used in parts (a) and (c). The larger magnetization precession amplitude observed for current pulses propagating along [110] direction is in agreement with the stronger $H_{SO}$ compared to the case of pulses propagating along [$\bar{1}$10] direction.

The MO pump-probe traces measured in the places indicated in Fig. 3(a) are shown in Fig. 3(c) using the corresponding colors. The curves were fitted by Eq. (1) and the extracted precessional amplitude $A$ is shown in Fig. 3(d), again using the same color-coding as in panels (a) and (c). Clearly, there is a considerable difference in the precession amplitudes for photocurrent pulses propagating along the [110] and [$\bar{1}$10] crystallographic directions. In part, this difference is due to a geometrical factor, since the magnetization, which is aligned along one of the easy axes [as indicated in Fig. 3(a)], has different angles with $H_{SO}$ generated by current pulses along [110] and [$\bar{1}$10] and the corresponding torque is proportional to $M \times H_{SO}$. However, taking this geometrical factor into account, the precession amplitude for current pulses propagating along the [110] direction is still about 40% larger than that for pulses propagating along the [$\bar{1}$10] direction. This difference cannot be explained by the Oersted magnetic field which is radially symmetric and, therefore, it has an equal amplitude for these two current propagation directions. On the other hand, the difference in the precession amplitudes is in agreement with the difference in the magnitudes of $H_{SO}$ depicted in Fig. 3(a), which is an additional independent experimental confirmation that the ultrashort SOT pulses contribute significantly to the excitation of the magnetization precession in our device.

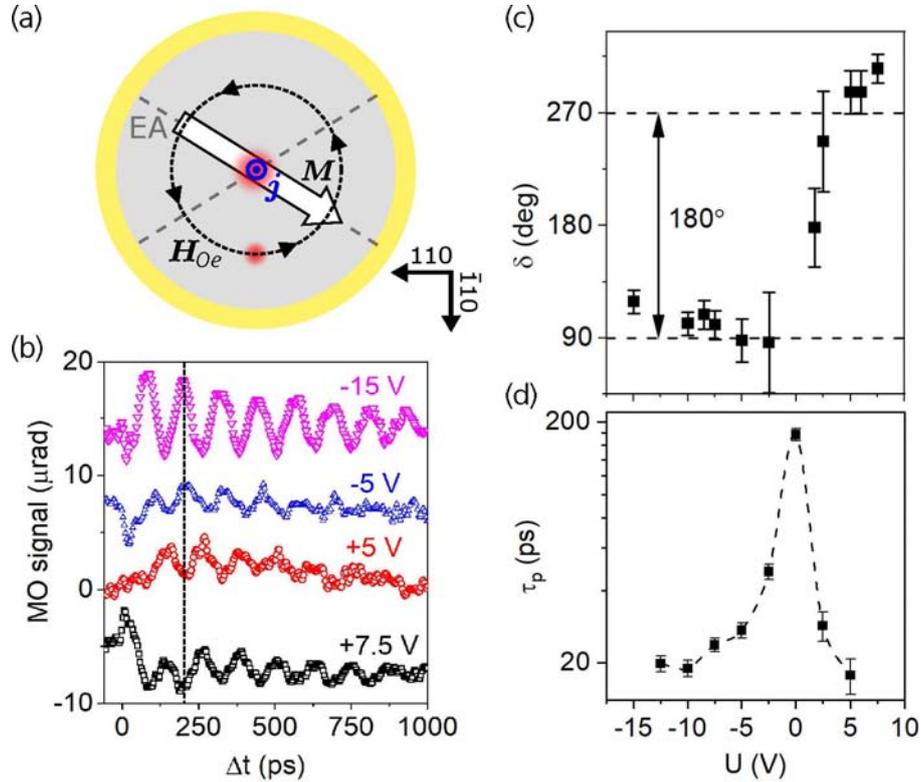

**Fig. 4 Magnetization precession generated by Oersted field pulses. (a)** Schematic illustration of the Oersted magnetic field $H_{Oe}$ (black circle) induced in the iron film by the vertically propagating current pulse, which is flowing in the pump-illuminated column, with a diameter of ≈ 1 μm, in the photodiode center (larger red circle). Switching the bias polarity reverses the photocurrent and, consequently, the corresponding field $H_{Oe}$ acting on magnetization $M$ within the probe laser spot (smaller red circle), which is displaced by 9 μm away from the pump laser spot along the [$\bar{1}$10] crystallographic direction. **(b)** Magnetization precession induced by $H_{Oe}$ measured for several biases without external magnetic field applied; the curves are vertically shifted for clarity. The phase of oscillations depends on the bias polarity, as indicated by the dashed line. **(c)** Bias dependence of the precessional phase $\delta$ extracted by fitting the data in part (b) by Eq. (1) for $\Delta t > 100$ ps. **(d)** Bias dependence of the duration of the vertically propagating photocurrent pulse measured by the photocurrent correlation technique. For larger bias magnitudes the photocurrent pulses become considerably shorter than the magnetization precession period, which is the cause of the gradual phase shift of the precession signal observed in (c).

To check the consistency of our analysis of the magnetization precession data and, in particular, the interpretation of the initial precession phase, we performed an additional control experiment in a geometry where the SOT magnitude is minimized. As shown in Fig. 3, this is the case when the current pulse propagates along the [$\bar{1}$10] direction. In these conditions, the Oersted field torque generated by the vertical current pulse becomes strongest relative to SOT generated by the lateral current pulse and, therefore, a precession phase closer to 90° is expected (see Supplementary Note 1 for details). The experimental geometry is depicted in Fig. 4(a). The probe laser spot is displaced along the [$\bar{1}$10] direction relative to the current pulse-generating pump laser spot that is placed in the center of the photodiode. The polarity of the photogenerated current pulse and, consequently, of the Oersted field ($H_{Oe}$) is determined by the polarity of the applied bias voltage. Moreover, the duration of the current pulse can be controlled by the bias magnitude (see Supplementary Note 5), which enabled us to verify experimentally its influence on the initial precession phase.

The pump-probe traces measured for bias voltages ranging from +7.5 V to -15 V are shown in Fig. 4(b). As indicated by the vertical dashed line, the magnetization precession changes phase by 180° when the bias polarity is reversed. This is expected, since an inversion of the current pulse propagation direction leads to an inversion of $H_{Oe}$ and the corresponding torque causes the opposite initial magnetization tilt. The initial precession phase $\delta$ extracted by fitting the curves with Eq. (1) is shown in Fig. 4(c) as a function of the applied bias voltage. In contrast to the conditions of Fig. 2, where the dominant contribution to torque was attributed to SOT, a phase of ≈ 90° is now observed for a small reverse bias voltage. This is fully consistent with the $H_{Oe}$–induced magnetization precession reported previously for a laser pulse-illuminated Schottky diode [9, 17, 18] and also with the phase theoretically predicted for an Oersted field pulse with a duration comparable to or longer than the precession period (see Supplementary Fig. S.1). This is indeed the case for small bias magnitudes in our diode, as is evident from Fig. 4(d) which shows the current pulse duration measured by the photocurrent correlation technique as a function of the applied bias voltage (see Supplementary Note 5). For larger bias magnitudes the precession phase shifts towards higher values for both bias polarities. This gradual phase shift, which is caused by the decreasing current pulse duration [see Fig. 4(d)], is in a quantitative agreement with the theoretically calculated phase dependence on pulse duration shown in Supplementary Fig. S.1.

## DISCUSSION

We have demonstrated generation of ultrashort spin-orbit torque pulses at an epitaxial ferromagnetic metal-semiconductor interface by converting femtosecond laser pulses into laterally-oriented high-amplitude current pulses in an electrically biased iron-gallium-arsenide (p-i-n) photodiode. Each ultrafast current pulse, consisting of a sub-picosecond peak and a ~ 10-picosecond tail, excites the precession of iron magnetization directly at the iron-photodiode interface without any dispersion-related pulse broadening. Polarity and lateral flow direction of the current pulses were controlled by the lateral position of a focused laser spot and by the bias voltage on the photodiode. The spin-orbit origin of the torque exciting Larmor precession of the iron magnetization was revealed by a detailed analysis of the precession phase and amplitude at different experimental conditions.

The experimentally observed Larmor precession of the ferromagnetic film with a frequency of ≈ 10 GHz was excited by the ~ 10-picosecond tail of the photogenerated current pulse. However, the much faster sub-picosecond current peak onset of the pulse with up to ~ 100 mA amplitude could be relevant for excitation of antiferromagnetic (AFM) thin films, where the exchange enhanced precession frequencies are in the THz regime [14]. Recent studies have demonstrated manipulation of the AFM Néel vector by passing an electric current through metallic collinear antiferromagnets with locally broken inversion symmetry, such as CuMnAs [19] or $Mn_2Au$ [20]. In particular, pulses with current densities of ~ $10^{10}$ - $10^{11}$ A/m$^2$ were sufficient to switch the Néel vector of the AFM domains by 180° due to domain wall motions [21, 22]. With our device structure, similar or even larger current pulse amplitudes could be generated within sub-ps time, providing a new experimental technique for investigation of ultrafast excitation and switching of magnetic order in metallic AFM thin films [23].

Moreover, as we demonstrate in Supplementary Note 2, thanks to the optical selection rules in the zinc-blende GaAs heterostructure, our new experimental technique is also able to exploit spin-orbit coupling to convert photon angular momentum to electron spin and, consequently, to generate ultrafast spin-polarized current pulses [24, 25, 26]. Direct spin transfer from ultrashort optically generated spin current pulse to the AFM layer might be another way of generation of high frequency AFM excitations.

**METHODS**

*Magneto-optical pump-probe experiment*

This technique was used to obtain information about the magnetization (spin) and charge dynamics that is excited in the photodiode directly by laser pulses or by the corresponding photocurrent pulses. Pump-probe experiment is a stroboscopic optical method where the dynamics triggered by a strong (pump) laser pulse is sampled by a time-delayed weaker (probe) laser pulse [27]. We employed a reflection optical geometry and measured the magneto-optical (MO) signal, which corresponds to a probe polarization rotation, and the differential reflectivity d$R$/$R$, which corresponds to a transient change of the probe intensity. Both these signals were measured simultaneously as a function of time delay between the pump and probe pulses using the optical bridge, where they correspond to "difference" and "sum" signals, respectively (see Appendix B in Ref. 28 for details). The measured MO signals did not depend on the polarization of pump pulses; the data depicted in Figs. 2, 3 and 4 were measured for linear polarization. As a light source, we used femtosecond Ti:sapphire laser (Mai Tai, Spectra Physics) producing ≈ 100 fs laser pulses at a repetition rate of 80 MHz. Typically, we employed a geometry with a collinearly propagating pump and probe pulses and a microscopic objective, which was focusing laser beams to a spot size of ≈ 1 μm; see Fig. 1(b) in Ref. 27. In the experiment, the laser repetition frequency was decreased to 8 MHz by a pulse picker, the laser central wavelength was tuned to 798 nm and spectral filters (NF808-34 in the pump beam and FBH810-10 in the probe beam) were used to generate a quasi-nondegenerate pump and probe pulses with central wavelengths of 794 nm and 804 nm, respectively [27]. We used pump laser power of 230 μW, which corresponds to a fluence of ≈ 1 mJ/cm$^2$, and probe laser power was set to 10 – 20% of the pump power. The data presented in Fig. 2, where a very good spatial overlap of the pump and probe pulses was necessary to suppress the influence of the Oersted field generated by a vertical photocurrent on the detected magnetization dynamics, were measured using an experimental setup with non-collinearly propagating pump and probe pulses where laser beams were focused by a single converging lens to a spot size of ≈ 25 μm; see Fig. 1(a) in Ref. 27. Here, the laser repetition rate of 80 MHz and the same wavelength of pump and probe pulses of 800 nm were used. The pump power of 60 mW, which corresponds to a fluence of ≈ 0.1 mJ/cm$^2$, and probe power of 10% of the pump power were used. The samples were mounted in an optical cryostat and the experiments were performed at a base temperature of 15 K. No external magnetic field was applied during the MO measurements.

*Photocurrent correlation method*

For the characterization of ultrashort electrical current pulses generated in our photodiode by femtosecond laser pulses, we used the photocurrent correlation measurement [29]. This technique is sensitive to the nonlinear component of the photodiode response, i.e., to the mutual interaction between two consecutive photo-generated current pulses [29, 30, 31]. The nonlinear response of our photodiode to the optical stimulus, which is necessary for this correlation technique to work [31], is clearly apparent from the VI characteristics shown in Supplementary Fig. S.4(b). Similarly like in the case of pump-probe method, we used two mutually time-delayed laser pulses which, however, now have the same intensity. Here, the measured quantity is a time-averaged photocurrent generated in the diode, which was measured by a sensitive ampere-meter as a function of the time delay between the two identical laser pulses. After the impact of the first laser pulse, the photo-generated electron-hole pairs are separated in a strong electric field present in the intrinsic part of the PIN diode and accelerated in the vertical direction towards the opposite electrodes, thus creating a short current pulse [see Fig. 1(a)]. The presence of the photo-carriers created by the first laser pulse influences both the material properties (e.g., via saturation of absorption) and the diode properties (e.g., by screening of the electric field) until the photo-carriers are removed from the diode by the electric field. The absorption of the second laser pulse is decreased because of the saturation and the electric field, which is weakened by the screening effect, makes the electrical transport slower, which means that more photo-carriers can recombine before reaching the electrodes. Both these effects decrease the total charge which can be transported out of the diode per one pair of laser pulses and, therefore, the ave Frage photocurrent measured in the correlation experiment is also decreased. As the influence of the first pulse on the second pulse increases with decreasing time delay between the pulses, the largest decrease of the average photocurrent is expected around the zero time delay.

**DATA AVAILABILITY**

The datasets generated and analyzed during the current study are available from the corresponding author on reasonable request.

**ACKNOWLEDGMENTS**

This work was supported in part by the Czech Science Foundation grant no. 19-28375X, by the EU FET Open RIA grant no. 766566, by the INTER-COST grant no. LTC20026, and by the European Union's Horizon 2020 research and innovation program under the Marie Skłodowska-Curie grant agreement No 861300 (COMRAD). We also acknowledge CzechNanoLab project LM2018110 funded by MEYS CR for the financial support of the measurements at LNSM Research Infrastructure and the COST Action CA17123 MAGNETOFON for the support of the international collaboration.

**AUTHOR CONTRIBUTIONS**

T.J., T.O., P.N., E.S., and J.W. designed the experiment and interpreted the data. T.J., T.O., P.N., and J.W. wrote the manuscript. T.J. and J.W. performed all optical and magneto-optical experiments. R.C., V.H., and V.N. grew the samples and Z.Š. performed the nano-lithography. T.O. carried out the theoretical modelling.

**ADDITIONAL INFORMATION**

*Competing interests:* The authors declare no competing interests.

# Ultrashort spin-orbit torque generated by femtosecond laser pulses
## Supplementary Material

T. Janda,[1,2,a)] T. Ostatnický,[1] P. Němec,[1] E. Schmoranzerová[1], R. Campion,[3] V. Hills,[3] Z. Šobáň,[4] and J. Wunderlich[2,4]

[1]*Faculty of Mathematics and Physics, Charles University, Ke Karlovu 3, 121 16 Prague 2, Czech Republic*

[2]*Institute for Experimental and Applied Physics, University of Regensburg, Universitätsstr. 31, 93053 Regensburg, Germany*

[3]*School of Physics and Astronomy, University of Nottingham, Nottingham, NG7 2RD, UK*

[4]*Institute of Physics ASCR, v.v.i., Cukrovarnická 10, 162 00 Prague 6, Czech Republic*

**Supplementary Note 1: Characteristic phase of magnetization precession excited in pump-probe experiment by various mechanisms**

The magnetization (*M*) precessional dynamics in ferromagnets is described by the Landau-Lifshitz-Gilbert (LLG) equation that can be expressed in a form [1, 2]:

$$\frac{d\boldsymbol{M}(t)}{dt} = -\mu_0\gamma\left[\boldsymbol{M}(t) \times \boldsymbol{H}_{eff}(t)\right] + \frac{\alpha}{M_s}\left[\boldsymbol{M}(t) \times \frac{d\boldsymbol{M}(t)}{dt}\right], \tag{S1}$$

where $\gamma$ is the gyromagnetic ratio, $\alpha$ is the Gilbert damping constant, and $M_s$ is saturation magnetization. The effective magnetic field $H_{eff}$ is defined as

$$\boldsymbol{H}_{eff}(t) = \frac{\partial \Phi}{\partial \boldsymbol{M}}, \tag{S2}$$

where $\Phi$ is energy density functional that includes Zeeman energy due to the external magnetic field $\boldsymbol{H}_{ext}$ and magnetic anisotropy energy consisting of the magnetocrystalline and shape anisotropy contributions. To describe the later discovered spintronic effects of spin-transfer and spin-orbit torques on the magnetization dynamics, an additional term $\boldsymbol{\tau}$ was added in Eq. (S1) resulting in the Landau-Lifshitz-Gilbert-Slonczewski (LLGS) equation [3]:

$$\frac{d\boldsymbol{M}(t)}{dt} = -\mu_0\gamma\left[\boldsymbol{M}(t) \times \boldsymbol{H}_{eff}(t)\right] + \frac{\alpha}{M_s}\left[\boldsymbol{M}(t) \times \frac{d\boldsymbol{M}(t)}{dt}\right] + \boldsymbol{\tau}(t). \tag{S3}$$

The first term on the right hand side of Eq. (S3) leads to a precession of magnetization about the direction of $\boldsymbol{H}_{eff}$ and the second term leads to relaxation of the magnetization towards this field. Consequently, depending on their symmetry, the spin torques can be sorted as field-like, of the form $\boldsymbol{\tau}_{FL} \sim \boldsymbol{M} \times \boldsymbol{s}$, and damping-like, of the form $\boldsymbol{\tau}_{DL} \sim \boldsymbol{M} \times (\boldsymbol{M} \times \boldsymbol{s})$ (see Fig. 3 in Ref.

3), with $s$ being the nonequilibrium spin polarization. The damping-like torque is also often called antidamping torque since, depending on the spin polarization direction, it can also turn magnetization away from $H_{eff}$.

After the impact of the pump laser pulse several effects take place, which can modify $H_{eff}$ and induce nonequilibrium spin polarization. Consequently, the corresponding torques $M \times H_{eff}$ and $\tau$ tilt the magnetization away from its equilibrium direction. The direction of this initial tilt influences the phase of the subsequent magnetization precession around the effective field. For example, the field-like and damping-like spin torques are always perpendicular to each other and, consequently, the phases of the magnetization precessions triggered by $\tau_{FL}$ and $\tau_{DL}$ differ by 90° (see Fig. S.1).

The magnetization precession phase is influenced not only by the particular excitation mechanism but also by the duration of the excitation pulse. This effect was discussed in Supplementary Note 1 in Ref. 4 for the extreme cases of very short and very long pulses (compared to the magnetization precession period). For a long pulse, the effective field is modified quasi-permanently and magnetization simply starts precessing around this new equilibrium direction. For a short pulse, the magnetization initially also starts precessing around the same new equilibrium direction. However, this initial stage lasts only for a time window that is much shorter than the magnetization precession period. During this time the magnetization is only slightly tilted and after the effective field is relaxed back to its original direction the magnetization starts its precession in a direction that is perpendicular to its initial tilt, i.e., perpendicular to the precessional motion triggered by the long pulse. Consequently, the magnetization precessions triggered by short and long pulses differ in phase by 90° (see Supplementary Fig. 1 in Ref. 4). The same considerations apply also to other excitation mechanisms that can be described by effective magnetic fields, such as the spin torques (see Supplementary Fig. 2 in Ref. 4).

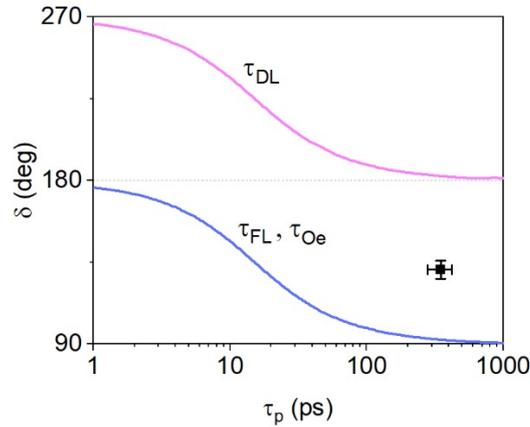

**Fig. S.1: Dependence of magnetization precession phase on the photocurrent pulse duration.** The lines depict magnetization precession phase $\delta$ deduced from simulated precessional dynamics triggered by Oersted-field pulses ($\tau_{Oe}$) and purely field-like ($\tau_{FL}$) and damping-like ($\tau_{DL}$) SOT pulses of different duration ($\tau_p$). Point: The experimentally measured phase and pulse duration corresponding to the experimental conditions of Fig. 2. A mixture of the damping-like and field-like SOTs is necessary to explain the observations.

In the following, we will review various physical phenomena that can be responsible for triggering magnetization precession after absorption of linearly polarized pump laser pulses in our GaAs-PIN/Fe structure and we will discuss how the different effects can be identified in the experimentally measured time-resolved magneto-optical (MO) data. Let us assume that the sample surface lies in the *xy*-plane, magnetization in equilibrium points in the *x*-direction and the pump-laser beam is incident on the sample along the *z*-direction. We also assume, in accord with our experimental conditions, that the experiment is performed in a reflection geometry and that the measured MO signal is due to polar Kerr effect. Consequently, the experimentally measured MO signal is proportional to perpendicular-to-plane component of magnetization $M_z$. By fitting the measured MO traces by a damped harmonic function in a form of Eq. (1) we can deduce the direction of the initial tilt of magnetization, which is one of the key signatures that can be used to identify the exact mechanism which was responsible for triggering the observed precession of magnetization [4, 5, 6, 7, 8, 9].

### *(a) Magnetic field pulse*

Seemingly, the most straightforward mechanism leading to a precession of magnetization is application of a magnetic field pulse with a sufficiently fast onset. When the pulse rise time is significantly shorter that the magnetization precession period, which is ≈ 100 ps in our Fe thin films, the magnetization can't adiabatically follow the effective field and starts precessing around it. As discussed in detail in Supplementary Notes 6 and 7, the current pulse generated in our Fe/GaAs structure, which consists of photogenerated electrons and holes, has two stages. Initially the electrons and holes move in vertical direction within the pump-illuminated channel and after reaching the top and bottom electrodes they spread in lateral direction towards the annular contact. The important difference is that during the vertical transport the electrons and holes move in opposite directions, while during the lateral motion they move in the same direction, coupled by the Coulomb interaction. Consequently, the Oersted magnetic fields generated by the electrons and holes during their vertical motion sum up, while the Oersted fields generated during their lateral motion subtract. Thanks to the large aspect ratio of our photodiode with the lateral size being 2 orders of magnitude larger than its thickness, the Oersted fields from the lateral electron and hole currents compensate almost precisely in the iron film and, therefore, they can be neglected.

The precession of magnetization induced by an in-plane Oersted magnetic field pulse, which results from a vertically propagating current pulse generated by a laser pulse excitation, has been previously studied in another vertical photoconductive structure – a Schottky diode [5, 6, 10, 11, 12]. The key fingerprint of this excitation mechanism is a characteristic dependence of the precession amplitude on the mutual spacing of the current pulse generating pump laser spot and the probe laser spot on the sample surface – the precession disappears when pump and probe pulses are overlapped spatially and it is the largest when there is some spacing between them (see Fig. 1(b) in Ref. 10). Furthermore, the resulting precession of magnetization is the largest when the Oersted field is perpendicular to the magnetization orientation and it disappears when these two vectors are parallel (see Fig. 2(a) in Ref. 10). This second feature is, however, true also for other effects, for example, for the current-induced spin-orbit torques discussed in part (c) of this Supplementary Note. Finally, since the Oersted field is oriented in-plane, the magnetization is initially tilted perpendicular to the sample plane, which for longer pulses leads to a precessional phase of 90° or 270°, depending on the mutual position of pump and probe laser spots and the magnetization direction (see Fig. 2 in Ref. 6). For pulses much shorter than the magnetization precession period a phase of 0° or 180° is expected.

### *(b) Thermal change of magnetic anisotropy*

As a matter of fact, the most frequently observed mechanism leading to a pump-induced precession of magnetization is a thermal mechanism, which appears due to different temperature dependencies of the various contributions to the effective field $H_{eff}$. The laser-pulse induced heating leads to a reduction of magnetization and, consequently, to a change in a relative strength of the various terms in $H_{eff}$, which scale differently with $M$. In particular, the Zeeman term is linear in $M$, while the magnetic anisotropy energy scales with higher powers of $M$ and, therefore, it is reduced relative to the Zeeman energy after the impact of the pump laser pulse. Consequently, when the external magnetic field $H_{ext}$ is applied at a certain angle with respect to the easy axis (EA) in the studied magnetic material, the laser pulse causes a reorientation of $H_{eff}$, which triggers the magnetization precession (see Fig. 1 in Ref. 13). All MO measurements presented in the main text were performed in zero external magnetic field and, therefore, this particular effect is not relevant in our case. Nevertheless, the thermal mechanism can trigger magnetization precession even in zero magnetic field – in materials, where magnetic anisotropy contains several contributions that have a different temperature dependence. For example, in ferromagnetic semiconductor (Ga,Mn)As the in-plane magnetic anisotropy consists of an uniaxial component, which scales with magnetization as $\sim M^2$, and a biaxial (cubic) component, which scales as $\sim M^4$. Due to the laser-induced heating, the uniaxial anisotropy is enhanced relative to the biaxial anisotropy, which leads to a reorientation of the EAs within the sample plane (see Fig. 3 in Ref. 14). The corresponding torque initially tilts the magnetization out of the sample plane and the subsequent precession around the modified EA continues for hundreds of picoseconds. Heat diffusion from the ferromagnetic film to the substrate eventually removes the excess heat and the original EA is restored. However, this relaxation usually proceeds at a timescale that is much longer than the magnetization precession period. Consequently, we expect an initial precession phase of 90° (or 270°). We note that our epitaxial Fe film also possesses a biaxial magnetic anisotropy composed from a uniaxial and a cubic component (see Supplementary Note 3 for details) and, therefore, this excitation mechanism might be relevant in our case.

### (c) Spin-orbit torque excitation mechanism

Spin-orbit torque (SOT) is generated by a nonequilibrium spin density which occurs due to the spin-orbit interaction. Typically, this phenomenon was studied in nonmagnetic metal (NM)/ferromagnetic metal (FM) bilayers. Here, in-plane charge current induces spin polarization mainly by two mechanisms [15, 16]: the bulk spin Hall effect (SHE) in NM layer and the interfacial inverse spin galvanic effect (iSGE). The SHE converts a charge current into a transverse spin current, inducing a spin accumulation at the NM/FM interface [15]. On the other hand, the iSGE generates a spin density due to spin-orbit coupling directly at the NM/FM interface, which lacks an inversion symmetry [15, 16]. In materials with bulk inversion asymmetry (like GaAs) the spin-orbit coupling has two distinct contributions known as the Rashba and Dresselhaus effects [17], which results in different strengths of SOT for currents flowing along different crystallographic directions [see Fig. 3(b)]. The nonequilibrium spin density $s$ accumulated at the NM/FM interface due to the spin-orbit effects acts as an effective spin-orbit field $H_{SO}$ which exerts a torque on the magnetization in FM. This torque can be decomposed into two parts: the field-like torque $\tau_{FL} \sim M \times H_{SO}$ and the damping-like torque $\tau_{DL} \sim M \times (M \times H_{SO})$. Since the spin-orbit field $H_{SO} \parallel s$ is oriented in-plane, $\tau_{FL}$ causes an initial tilt of the magnetization perpendicular to the sample plane, whereas $\tau_{DL}$ acts within the sample plane [see Fig. 2(a)]. Consequently, for $\tau_{FL}$ we expect the magnetization to precess with an initial phase of 90° (or 270°) and 0° (or 180°) for long and short current pulses, respectively. For $\tau_{DL}$, which is perpendicular to $\tau_{FL}$, the phases will be correspondingly shifted by 90°.

SOT is far less understood than the other mechanisms. It is typically used to electrically manipulate the magnetization in memory devices [15], but it is usually limited to nanosecond timescales. Only very recently, the switching with picosecond-long current pulses due to SOT

## (d) Identification of the excitation mechanisms responsible for the magnetization precession induced by photocurrent pulses in our Fe/GaAs PIN-diode

First of all, in the experimental geometry of Fig. 2, i.e., with the spatially overlapped pump and probe laser pulses, there is no net effect from the Oersted field generated by the vertically propagating current pulse (see Fig. 1(b) in Ref. 10). Furthermore, the absence of the magnetization precession for the case of the laser spots overlapped in the device center [position (ii) in Fig. 2(b)] and the opposite phases of the precessions excited in positions (i) and (iii), as well as the bias dependence of the precessional phase in Fig. 4(c), rule out the thermal change of magnetic anisotropy as a dominant mechanism triggering the precession. Consequently, what remains are only excitation mechanisms that are connected with the generated *lateral* photocurrent pulses. As already mentioned, the effect of the corresponding Oersted field can be neglected due to the compensation of the fields from the electron and hole lateral currents flowing in the top and bottom electrode, respectively. We can also exclude effects due to Joule heating caused by the photocurrent pulses which would be independent of the current direction, i.e., which would lead to identical precessional phases in positions (i) and (iii) in Fig. 2(c). Therefore, the only remaining excitation mechanism which can appear in the experimental conditions of Fig. 2 and which is compatible with our observations is the spin-orbit torque.

The presence of SOT was further confirmed experimentally by an analysis of the induced magnetization precession. There are two characteristic signatures of SOT that can be identified in the precession phase and amplitude. First, the precession phase of $\approx 135°$ observed in Fig. 2 can be explained neither by the thermal mechanism nor by the Oersted-field pulse, which would both lead to a precession phase of $90°$ for current pulses longer than the precession period (see Supplementary Note 5 for the measurements of the current pulse duration). On the other hand, the observed phase can be well explained by SOT, in particular, by the presence of damping-like SOT. As already mentioned, the field-like and damping-like SOTs would induce precessions with phases of $90°$ and $180°$, respectively. Therefore, if both contributions are present, one would observe a phase in the interval of $90°$ - $180°$, depending on the ratio of their amplitudes. This is exactly what we observe in Fig. 2(d), indicating a strong contribution of SOT to the excitation of the magnetization precession. Second, the combination of Rashba and Dresselhaus spin-orbit fields leads to different magnitudes of SOT for currents along different crystallographic directions and, consequently, to differences in the amplitudes of the excited magnetization precessions. This asymmetry is apparent in measurements shown in Fig. 3, where we controlled the direction of the current pulse propagation by positioning the probe laser spot with respect to the pump spot. The observed larger precessional amplitude for current pulses along the [110] direction is in agreement with the larger $\boldsymbol{H}_{SO}$ present in this case, compared to pulses along the [$\bar{1}$10] direction. Both arguments, based on the precessional phase and amplitude, prove a significant contribution of the ultrashort SOT pulses to the excitation of Larmor precession in our device.

Finally, we will discuss the influence of the current pulse duration on the observed magnetization precession phase. So far, we only considered pulses much shorter or longer than the precession period, for which the initial phase differs by $90°$. Obviously, pulse durations in between these two extreme cases lead to some intermediate values of the phase. We performed a simulation of the magnetization precession excited by a SOT pulse with an instantaneous onset and mono-exponential decay time $\tau_p$. In Fig. S.1 we show the precessional phase δ as a function of $\tau_p$ for the case of a sole action of $\boldsymbol{\tau}_{FL}$ and of $\boldsymbol{\tau}_{DL}$. We note that for an action of the Oersted-field-induced torque $\boldsymbol{\tau}_{Oe}$, which is present in the measurements with spatially separated pump and probe laser spots (Figs. 3 and 4), the results are identical as for $\boldsymbol{\tau}_{FL}$. The phases were obtained from the simulated dynamics by fitting the oscillations with a cosine function for times > $5\tau_p$, when the magnetization is already precessing around the original unperturbed effective field. The simulation confirms the conclusions of parts (a) and (c) regarding the precessional phases that are expected for Oersted-field and SOT pulses much longer / shorter than the magnetization precession period of ≈ 100 ps. In particular, significant phase modification occurs when the pulse duration drops down to ≈ 20 ps, which is in a perfect agreement with our observations of the gradual phase shift for large bias magnitudes in the experiment with ≈ 1 μm-wide spatially-separated laser spots [cf. Figs. 4(c) and 4(d)]. On the other hand, the precessional phase of ≈ 135° observed with the spatially-overlapped ≈ 25 μm-wide laser spots [see Fig. 2(d)] cannot be ascribed to the influence of the current pulse duration, since $\tau_p$ in this case is ≳ 300 ps (see Supplementary Note 5 for details). Consequently, the observed precessional phase can only be explained by an action of SOT, in particular, by a mixture of damping-like and field-like SOTs, as indicated by the experimental data point in Fig. S.1, which represents the phase and pulse duration corresponding to the experimental conditions of Fig. 2.

**Supplementary Note 2: Generation of spin-polarized current pulses**

The ultrashort current pulses generated in our photodiode can be spin-polarized thanks to the specific band structure of the used semiconductor. Spin-polarized electrons and holes are generated in GaAs by absorption of circularly polarized light due to the effect of optical orientation [19]. The direction of the photoinjected spins is controlled by the direction of light propagation and by its helicity [20]. While the photo-holes lose their spin polarization in less than 100 fs [21], the photo-electrons can maintain their spin polarization for several nanoseconds in optimally *n*-doped semiconductors [22, 23] and, therefore, can exert a torque on magnetization. This was experimentally demonstrated in a ferromagnetic semiconductor GaMnAs [24], where the spin-polarized electrons were generated directly in the magnetic material.

Our device design, where a vertical photodiode structure made of GaAs semiconductor is combined with a thin metallic ferromagnetic layer, offers a possibility to inject ultrashort spin-polarized current pulses also to metals where spins can't be generated directly by light absorption. Under a reverse bias the photo-electrons generated in the intrinsic layer of the diode are accelerated towards the magnetic layer, which is deposited directly on top of the diode, creating an ultrashort spin-polarized current pulse with a sub-picosecond onset. We note that this method of generation of ultrashort spin-polarized electrical pulses using circularly polarized laser pulses can be applied neither in a standard lateral photoconductive switch (Auston switch) [25, 26] nor in a Schottky diode [10], which have been previously used to optically generate picosecond current pulses. In the Auston switch the active area is formed by a gap in a metal electrode, which is deposited on top of a semi-insulating semiconductor substrate, where electric bias is applied. Illumination of the gap by a femtosecond laser pulse (with an appropriate wavelength) generates electron-hole pairs in the substrate which enable temporarily a current to flow across the gap. In the case of circularly polarized laser pulses the corresponding current pulses can be spin-polarized. The photogenerated current pulses are then delivered to the investigated magnetic layer via an attached coplanar waveguide, typically over a distance of several tens of micrometers [18, 27, 28, 29]. However, the spin polarization of the photo-electrons is lost during the transport in the metal waveguide already after tens of nanometers [4, 30]. This spatial separation of the injection point of spin-polarized carriers and the position of investigated magnetic layer is not an issue in the Schottky diode, where the metal layer is deposited directly on top of the semiconductor. Nevertheless, in the typically used Schottky barrier formed between the *n*-type semiconductor and the metal, the built-in electric field accelerates the spin-polarized electrons away from the (magnetic) metal (see Fig. 3 in Ref. 10). Here, the charge current pulse is transported by holes, which are not spin-polarized due to their ultrafast (sub-100 fs) spin relaxation time [21].

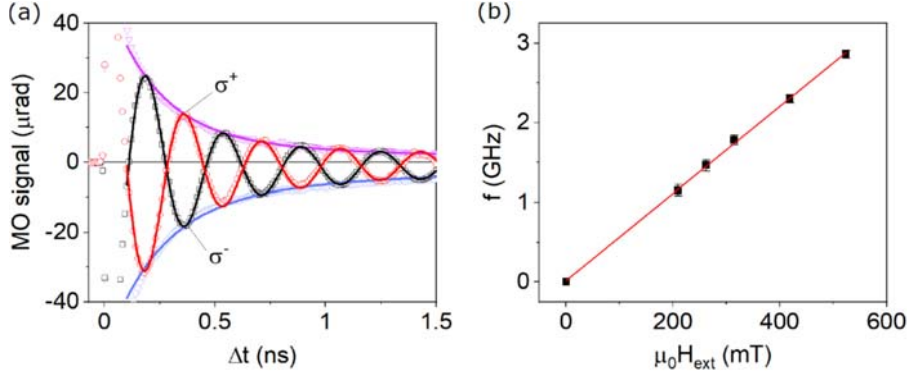

**Fig. S.2: Electron spin dynamics in the Fe/GaAs PIN-diode structure. (a)** Precession of the photo-injected spin-polarized electrons measured in perpendicularly applied external magnetic field of 500 mT (oscillatory signals) and without the field (curves with monotonous decays) for $\sigma^+$ and $\sigma^-$ circularly-polarized laser pulses (points). The lines are fits by a double-exponential decay function with time constants $t_1 = 240 \pm 10$ ps and $T_2^* = 1470 \pm 90$ ps. **(b)** Electron Larmor precession frequency as a function of the external magnetic field (points). The line shows the fitted linear dependence with electron g-factor of $0.39 \pm 0.02$.

The long-lived spin polarization of electrons photo-generated in our device was confirmed by time-resolved MO measurements performed using the pump-probe setup (see Methods section in the Main text). In Fig. S.2(a) we show the MO signals measured for opposite helicities of the pump laser pulses both with and without external magnetic field $\boldsymbol{H}_{ext}$. For $\boldsymbol{H}_{ext}$ applied in the sample plane, i.e., perpendicular to the photoinjected spin, the spins are forced to precess around $\boldsymbol{H}_{ext}$. Consequently, the measured MO signal, which is proportional to the spin polarization projection to the probe beam propagation direction, gives rise to the oscillatory signals. For opposite laser helicities the curves are phase-shifted by 180° due to the opposite directions of the spin at the instant of photo-generation. For $H_{ext} = 0$ the electron spin only decays without precession, which manifests as the envelopes of the oscillating curves in Fig. S.2(a). From the measured linear dependence of the oscillatory (Larmor) frequency on $H_{ext}$ [see Fig. S.2(b)] we inferred the magnitude of the g-factor of $0.39 \pm 0.02$, which confirms that the spin carriers are electrons in GaAs [31]. The measured damping of the MO signal can be fitted by a double-exponential decay where the shorter and the longer time constants can be attributed to the lifetime of photoinjected carriers and to the electron transverse spin coherence time $T_2^*$, respectively (see Eq. (1) in Ref. 23 and the adjacent discussion). From the context of this paper, the most important observation is that the measured electron spin coherence time $T_2^* \sim 1.5$ ns is considerably longer than the duration of electrical pulses (see Fig. S.5), i.e., that the photoinjected electrons are spin polarized in GaAs when they contribute to the electrical pulse.

In the experiment reported in the Main paper, the iron magnetization precession did not depend on the polarization of pump laser pulses. The absence of pump helicity-dependent oscillatory MO signal indicates that the spin polarization of the current pulses was not sufficient to overcome the large demagnetizing field of the iron film, which acts against the spin transfer torque-induced perpendicular-to-plane tilt of the magnetization, which is triggering the precession [8, 24, 32, 33]. Alternatively, the spin polarization of the current pulses might have been reduced while crossing the interface between GaAs and Fe making the efficiency of the spin transfer torque too low to be experimentally observable.

**Supplementary Note 3: Magnetic anisotropy of Fe/GaAs epilayer**

The studied sample contains iron film with a thickness of 2 nm, which corresponds to approx. 14 monolayers (ML) of Fe, that is grown on GaAs(001). Magnetization in the sample lies in the film plane due to a strong (≈ 2 T) demagnetizing field. The in-plane magnetic anisotropy of the film was characterized by a SQUID magnetometer. The $M(H_{ext})$ dependencies measured along three different crystallographic directions are shown in Fig. S.3(a). By fitting the data to the theoretical expressions given in Ref. 34 we identified an uniaxial anisotropy component with easy axis (EA) along the [110] direction and anisotropy field of $\mu_0 H_u = -13 \pm 3$ mT, and a cubic component with EAs along [100] and [010] directions and anisotropy field of $\mu_0 H_c = 30 \pm 6$ mT. This magnetic anisotropy is quite unusual for such an ultrathin film because in very thin films (≲ 25 ML) the uniaxial anisotropy typically dominates [35]. A possible explanation is the influence of interfaces. It is known that the cubic anisotropy has both surface and volume contributions while the uniaxial one originates purely from the interfaces (both Fe/GaAs and Fe/cap). These surface anisotropies are very sensitive to a surface quality, termination of the substrate, material of the capping layer or eventual mixing of Fe with the materials at the interfaces. These surface effects become increasingly important with decreasing the thickness of the Fe layer and can significantly modify the overall magnetic anisotropy in very thin films. Presumably, they are responsible for the observed magnetic anisotropy in our sample.

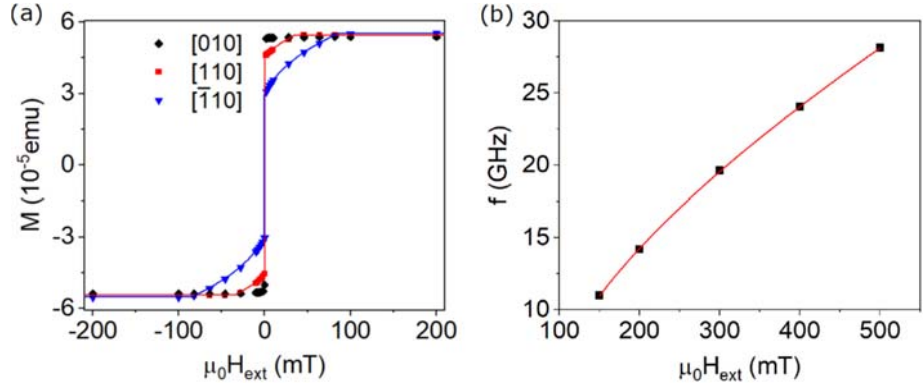

**Fig. S.3: Magnetic characterization of the Fe/GaAs epilayer. (a)** Magnetization components measured by SQUID magnetometry along different crystallographic directions. **(b)** Dependence of magnetization precession frequency *f* measured in a pump-probe MO experiment as a function of external magnetic field $H_{ext}$ applied along the [$\bar{1}$10] direction (points). The line is a fit by Eq. (S5).

To verify independently the deduced magnetic anisotropy, we measured in magneto-optical pump-probe experiment a dependence of magnetization precession frequency on external magnetic field applied along the [$\bar{1}$10] direction. For fields larger than ≈ 150 mT, the magnetization points along the field direction and the precessional frequency follows the Kittel formula [36]

$$f = \frac{\mu_0 \gamma}{2\pi} \sqrt{(H_{ext} + M_S^{eff} + H_C + 2H_U)(H_{ext} - 2H_C + 2H_U)}, \qquad (S4)$$

where $\gamma = (g\mu_B)/\hbar$ is the gyromagnetic ratio, $g$ is the Landé g-factor, $\mu_B$ is the Bohr magneton, and $\hbar$ is the reduced Planck constant. $H_{ext}$ is the magnitude of the external magnetic field, $H_c$ and $H_u$ are the anisotropy fields corresponding to the in-plane cubic and uniaxial contributions, respectively. The effective magnetization $M_S^{eff} = M_S - 2H_{out}$ contains the contributions from the shape anisotropy (demagnetizing field), which is connected with the saturation magnetization $M_s$, and the out-of-plane uniaxial anisotropy $H_{out}$. While in the first bracket in Eq. (S4) the material anisotropy fields do not play an important role (as it is dominated by the demagnetizing field $\mu_0 M_s \approx 2$ T), in the second bracket in Eq. (S4) the magnitude of $H_c$ and $H_u$ have a significant influence. Therefore, we define an effective anisotropy field $H_a^{eff} = H_c - H_u$ and the fitting formula reduces to

$$f = \frac{\mu_0 \gamma}{2\pi} \sqrt{(H_{ext} + M_S^{eff} + H_a^{eff} + 3H_U)(H_{ext} - 2H_a^{eff})}. \tag{S5}$$

Because the fitting procedure is almost insensitive to the precise value of $H_u$ in the first bracket in Eq. (S5), we set there $\mu_0 H_u = -13$ mT, as measured by SQUID. The measured dependence of precession frequency on $H_{ext}$ is shown in Fig. S.3(b) together with the corresponding fit. The obtained g-factor of 2.1 exactly agrees with the value reported in literature for Fe [34], confirming that the oscillatory MO signal corresponds to the precession of magnetization in iron. The effective anisotropy field $\mu_0 H_a^{eff} = 36 \pm 4$ mT also agrees very well with the value obtained by SQUID ($\mu_0 H_c$ - $\mu_0 H_u = 43 \pm 9$ mT). Finally, the obtained effective magnetization $\mu_0 M_S^{eff} = 1.67 \pm 0.25$ T gives for a typical value of $\mu_0 H_{out} = 235$ mT [35] a value $\mu_0 M_s = 2.14$ T, which perfectly agrees with the saturation magnetization of bulk Fe [35].

**Supplementary Note 4: Electrical properties of Fe/GaAs photodiode**

The V-I characteristic of our photodiode measured in darkness is shown in Fig. S.4(a). The diode can withstand large bias magnitudes of both polarities, unlike previously used Schottky diodes [10, 11]. The ability to reverse the direction of electric field in the intrinsic layer of our diode and, consequently, the direction of the photocurrent pulse propagation was important to identify the effects of the current-induced torques on iron magnetization.

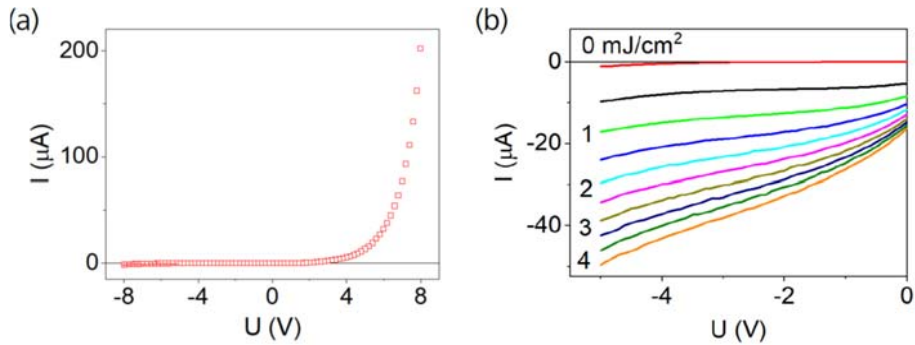

**Fig. S.4: Electrical characterization of the Fe/GaAs photodiode. (a)** V-I characteristics of the photodiode in darkness showing the ability to withstand large biases of both polarities. **(b)** V-I characteristics under illumination with a single fs-laser beam showing a nonlinear dependence of the photocurrent on the laser fluence.

The behavior of the photodiode under illumination is shown in Fig. S.4(b). The individual V-I curves were measured using a single fs-laser beam focused to the device center with the laser fluency increasing in constant steps of 0.45 mJ/cm2. Apparently, the measured photocurrent increases sub-linearly, which indicates a decreasing charge collection efficiency with an increasing photo-carrier density. This is caused mainly by the interplay between the internal electric field and the photo-generated charge [37, 38], as explained in detail in Supplementary Note 6. This non-linear behavior is necessary for the photocurrent correlation technique to work.

**Supplementary Note 5: Characterization of the electrical pulses by photocurrent correlation technique**

As mentioned in the Main text, the later stages of the current pulse decay in the presence of the screening electric field were investigated by the photocurrent correlation technique (see the Methods section). In Fig. S.5(a) we show the correlation curves measured with the laser spots having a diameter of ≈ 1 μm, which corresponds to the experimental conditions of Figs. 3 and 4 in the Main text. The measured photocurrent is plotted relative to its value at zero time delay for different external biases of both forward and reverse polarities. As expected, the photocurrent is reduced around $\Delta t = 0$ (see Methods for details). Note that in our convention, the current is negative for reverse (negative) bias and positive for forward (positive) bias, i.e., in both cases the photocurrent magnitude decreases for small time delays.

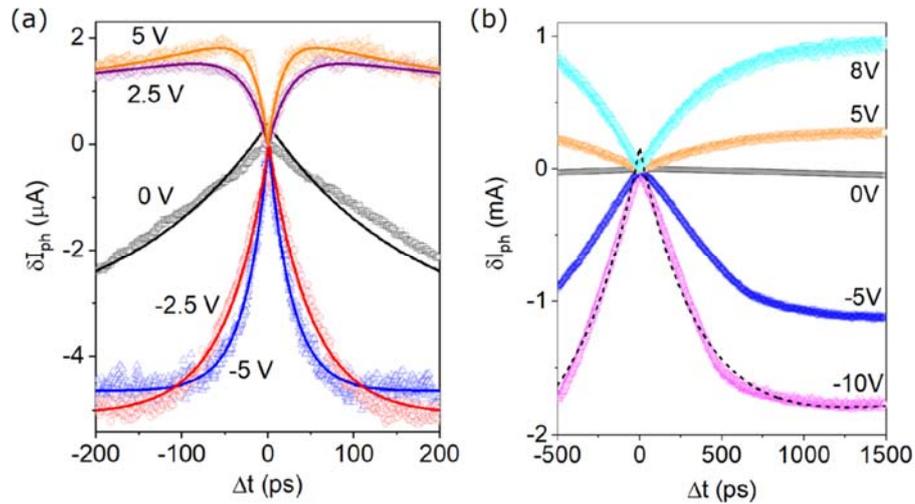

**Fig. S.5: Characterization of ultrashort photocurrent pulses by photocurrent correlation technique. (a)** Correlation measurements performed with ≈ 1 μm-wide laser spots. The time-averaged photocurrent $I_{ph}$ is measured as a function of the mutual time delay $\Delta t$ between two identical laser pulses for different applied biases. The change of photocurrent relative to its value at zero time delay, $\delta I_{ph} = I_{ph} - I_{ph}(\Delta t=0)$, is plotted. The measured data (points) are fitted (lines) by a single- and double-exponential model for reverse and forward biases, respectively. **(b)** Analogous measurements performed with ≈ 25 μm-wide laser spots (points) showing substantially longer current pulses due to more pronounced electric screening effects. Dashed line: fit by a single-exponential model.

The interpretation of the measured correlation signal is not straightforward. The theoretical analyses performed in Refs. 38 and 39 link the temporal profile of a single photocurrent pulse to the profile of the measured time-averaged correlation signal. In their case of a lateral photoconductive switch, the double-exponential profile of the correlation signal was attributed to the exponential decay of the photocarriers (and the corresponding photocurrent) via trapping and recombination. As shown in Fig. S.5(a), the photocurrent correlation data from our PIN diode can be fitted well by a single-exponential model

$$I_{ph}(\Delta t) = I_0 + I_p\, e^{-\frac{\Delta t}{\tau_p}} \tag{S6}$$

for larger reverse biases. Here, $I_0$ is an offset that corresponds to a dark current plus the photocurrent for well time-separated laser pulses, $I_p$ is the photocurrent drop induced by overlapping the two current pulses in time, i.e., it is a measure of the nonlinearity of the system, and $\tau_p$ corresponds to the decay time of the photocurrent pulse. For a low or zero bias, the electric field screening is more pronounced, which might explain the slightly distorted correlation curve. However, based on the aforementioned theoretical analyses, we assume that also in our case the decay time of the photocurrent correlation reflects the overall duration of the underlying ultrashort photocurrent pulses. This was additionally supported by the performed numerical simulations (see Supplementary Note 6 for details). In the case of a forward bias, the photocurrent correlation obviously contains an additional component with an opposite polarity, i.e., a component which corresponds to an effect increasing the photocurrent with decreasing the time delay. We ascribe this contribution to the effect of heating. Each laser pulse absorbed in the diode heats the semiconductor, which temporarily increases its conductivity and, therefore, the forward current supplied by the external electrical source (we apply a constant voltage). Since the conductivity of an intrinsic semiconductor increases super-linearly with temperature, $\sigma \sim \exp(-E_g/(2k_BT))$, the closer the two laser pulses are in time, the larger is the temperature increase and correspondingly larger is the average current flowing through the diode. We took into account the heating effect by adding a second exponential term, $\exp(-|\Delta t|/\tau_h)$, to Eq. (S6), where $\tau_h$ corresponds to the time-scale at which the pulse-induced heat is dissipated from the illuminated spot. The decay times $\tau_p$ determined by the fitting procedure are shown in Fig. 4(d) in the Main text as a function of the applied bias. Clearly, the decay time decreases rapidly with increasing bias and saturates at $\approx 25$ ps for biases with magnitude larger than 5 V. This is expected, as the stronger electric field depletes the photo-generated charge faster. From the fitting we obtained also the heating-related time constant $\tau_h = 210 \pm 30$ ps. We note that the deduced decay time is limited mainly by a transport of the screening charge from the central area of the device to the circular draining electrode (see Supplementary Note 6 for a detailed discussion). Therefore, shorter and stronger current pulses can be expected in devices of smaller diameter.

In Fig. S.5(b) we show the photocurrent correlation curves obtained with laser spots having a diameter of $\approx 25$ μm, which corresponds to the experimental conditions of Fig. 2 in the Main text. Apparently, the correlation signal decays much slower than in the case of the smaller laser spots. We fitted the correlation curve obtained for a bias of -10 V by Eq. (S6) supplemented by an additional linear function to account for the linear slope. This linear background was probably caused by the fact that the two time-delayed laser pulses were not absolutely identical or by slight changes in the properties of one of the laser spots when the time delay was changed by the optical delay line, which only become noticeable for a large range of time delays. The exponential decay time obtained by fitting of $\approx 350 \pm 70$ ps is an order of magnitude longer than that for current pulses generated by the small laser spots. It is also significantly longer than the magnetization precession time and, therefore, the measurements with the overlapped pump and probe laser spots shown in Fig. 2 were performed in the limit of a "long" current pulse, in the context of the discussion in Supplementary Note 1. The longer duration of the current pulses generated by the 25 μm-wide laser spot is caused by much larger amount of photogenerated charge and, consequently, more pronounced screening effects that prevent a fast drainage of the photocarriers from the device. The large amount of photogenerated charge is also apparent from the measured time-averaged photocurrent magnitudes of several mA compared to tens of μA in the case of 1 μm-wide laser spots.

**Supplementary Note 6: Theoretical modelling of photocurrent pulses**

The system under consideration is in the initial state composed of electron-hole pairs at low temperature. The calculated density of photo-excited pairs is below the exciton Mott density and the thermal energy is below the exciton binding energy. The electric field which is a sum of the external and the built-in fields is, however, large enough to prevent electron-hole binding so the system may be regarded fully as a two-component classical plasma without quantum correlations caused by the electron-hole Coulomb interaction. The only microscopic effect, reflected in the systems dynamics, is then the electron-hole recombination.

We calculated the initial distribution of the electron-hole density, considering the propagation of a laser pulse at a given wavelength within a saturable absorbing medium. The saturation density of electron-hole pairs was set to $3\times10^{17}$ cm$^{-3}$, according to a microscopic estimate based on a three-band model of *i*-GaAs. The subsequent dynamics of the two-component plasma was simulated with a FDTD algorithm using a two-dimensional mesh. Thanks to the rotational symmetry, all variables are constant over the angular coordinate in the cylindrical coordinate system and, therefore, they depend only upon the longitudinal and radial positions, which are denoted hereafter as vertical and lateral, respectively – see Fig. 1(a). The *p*- and *n*-doped layers were considered as two single layers between whom the *i*-GaAs area was sandwiched. The topmost ultrathin (2 nm) Fe layer was not included as a separate layer in the simulations because its effect on the overall systems dynamics is not considerable. The reason is that the magnitude of the lateral current is in our particular geometry determined solely by the small conductivity of the bottom *p*-doped layer due to the attractive Coulomb force between laterally propagating electrons and holes. An effective increase of the conductivity of the topmost layer has, therefore, no effect at all.

We used the following parameters in the simulations: the device radius of 50 μm was divided to 256 cells, the i-GaAs layer thickness of 1 μm was divided to 24 cells with the electron and hole mobilities 20 000 cm$^2$/Vs and 1 000 cm$^2$/Vs, respectively. For *p*-GaAs the layer thickness, majority charge concentration and its mobility were 500 nm, $2\times10^{18}$ cm$^{-3}$ and 200 cm$^2$/Vs, respectively. Finally, the parameters for the *n*-GaAs layer were 20 nm, $5\times10^{18}$ cm$^{-3}$ and 2 000 cm$^2$/Vs [40, 41]. The electron-hole recombination time considered in our simulations was 100 ps. The device with two electrodes under the reverse bias may be regarded as a charged capacitor with an additional built-in field due to the *p-i-n* junction. The photo-injected charges generated between the electrodes spatially separate to a positively and a negatively charged cloud, each of them traveling towards the oppositely charged electrode (i.e., electrons in the direction of the *n*-GaAs for the reverse bias). This motion results in a screening of the internal field and in slowing down the motion of the screened charges. The charge which eventually hits the electrodes discharges the capacitor and is quickly drained to the electric circuit due to the external voltage, thus rebuilding the internal field and transferring the rest of the charge towards the electrodes. In our device, however, the draining of the charges from electrodes is not immediate as they need to travel from the illuminated center of the disc to its edge to reach the draining contact. The trajectory of each electron (hole) can be, therefore, separated to the part where it propagates vertically towards one of the electrodes and a subsequent lateral motion to the edge of the structure – see Fig. 1(a). This separation of current to two distinct routes, which is one of the major assets in our device structure, is justified by the proportions of the system: ratio of the radius to the height is 50:1 and also the width of the excitation spot is much smaller than the device radius. The vertical electric field, even when screened by the charges in the electrodes, therefore represents the major driving force and the horizontal electrical field becomes important only when the vertical one vanishes.

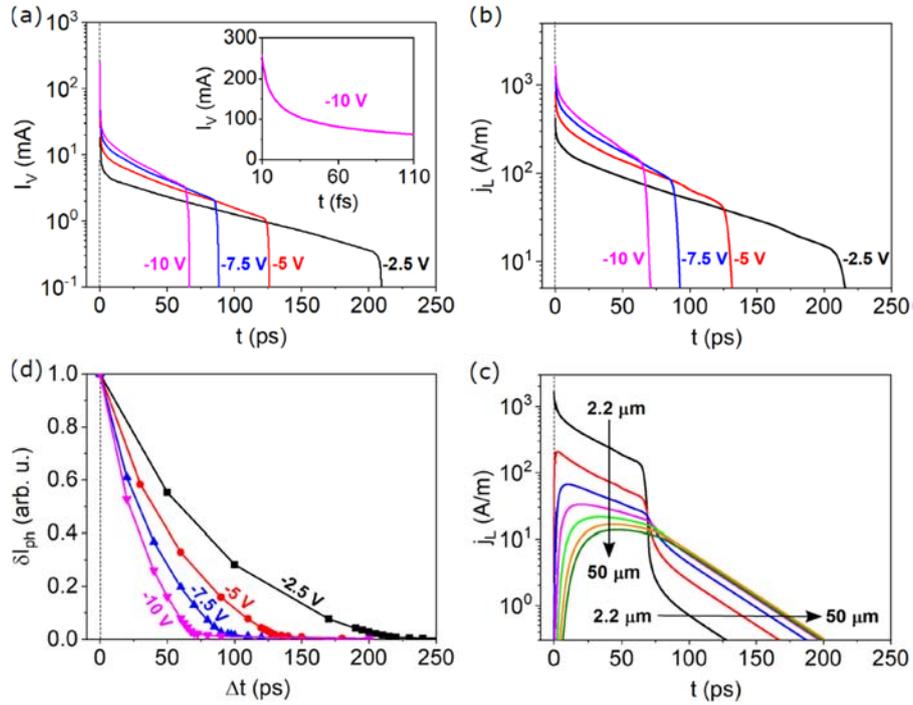

**Fig. S.6: Numerical simulations of the charge transport dynamics.** For different reverse voltage biases we show time evolutions of: (a) vertical current entering the *n*-electrode, (b) lateral current density flowing in the *n*-electrode at the edge of the illuminated area, and (d) corresponding normalized photocurrent correlation curves. The inset in part (a) depicts the very rapid initial dynamics of vertical current for -10 V, assuming that charge carriers are photoinjected by a $\delta$-pulse (see Supplementary Note 7 for details). (c) Lateral current density time profiles at different distances (2.2, 8, 16, 25, 33, 42, and 50 μm) from the center of the illuminated spot for the applied voltage of -10 V. Note that y-scales in parts (a)-(c) are shown in logarithmic scale. As we discuss in Supplementary Note 7, to highlight the principal application potential of our device structure, we assumed in our simulations that charge carries are photoinjected by a $\delta$-pulse. In reality, the excitation laser pulse has a Gaussian time profile with a FWHM of 100 fs and, consequently, the realistic current time profiles are convolutions of the laser pulse and the profiles shown in this Figure.

Results of the numerical calculations are shown in Fig. S.6. The distinct stages of the systems evolution can be well recognized in Fig. S.6(a) where we plot the evolution of the vertical electron current. In the initial phase, there is a large current immediately after the photo-generation of charge carriers which decreases non-exponentially due to a fast buildup of a screening in the electrodes. Then, after several picoseconds, the screening charge slowly leaves the photoexcited central area of the device while it is being instantly refilled by the vertical current. Finally, after all the charge is drained from and/or recombines in the *i*-GaAs area, the vertical current ceases. In Fig. S.6(b) the magnitude of the lateral current at the edge of the photoexcited area reveals exactly the same behavior since once the charges enter the electrode, they migrate towards the edge of the device. The only difference is that there is a residual current due to the screening charge after the cease of the vertical current. The lateral current pulse spreads in time while propagating towards the drain contact – this feature is depicted in Fig. S.6(c) where we observe the initial fast ($\approx$ 50 ps) current peak only at positions not further than 25 μm from the device center. Closer to the device edge, the current pulse is characterized by a slow initial build-up and then a monoexponential decay. The simulation of the photocurrent correlation measurement is shown in Fig. S.6(d). As already described, its initial fade-out is an effect of the removal of the electric charge from the optically active region, which is driven by the recombination plus its transport to the electrodes and further to the drain contacts. The time constant of the exponential fit of the correlation functions, which is shown in Fig. 4(d) as a characterization parameter describing the experimentally measured photocurrent correlation curves, gives an estimate of the overall length of the vertical current pulse [cf. Figs. S.6(a) and S.6(d)]. Note, however, that – unlike in transient reflectivity measurements [see Fig. 1(c)] – the initial ultrafast stage of the vertical current [see the Inset in Fig. S.6(a)] is not apparent in the correlation function due to rather different dynamics of electrons and holes. Further theoretical analysis of the properties of generated electrical pulses and resulting magnetic and SOT pulses, including their optimizations by changing the device structure and dimensions and/or properties of the excitation laser pulses, are provided in Supplementary Note 7.

**Supplementary Note 7: Theoretical predictions about possible future device optimizations**

In the current dynamics of the studied device, there are several time constants of interest. Here we discuss them separately with a focus on eventual future experiment optimizations by changing the device structure and dimensions and/or properties of the excitation laser pulses. In the discussion we consider only the consequences for the Oersted-field pulses generated by the vertical current pulses. However, very similar conclusions apply also to the SOT pulses induced by the lateral current pulses, as both the Oersted field and $H_{SO}$ are proportional to the current density and the temporal profile of the lateral current pulse at the edge of the illuminated area is very similar to that of the vertical current pulse [cf. Figs. S.6(a) and S.6(b)].

*(a) Initial fast decay of vertical current due to electric field screening*

As discussed in Supplementary Note 6, the electron-hole cloud is created within the electric potential between charged electrodes of the *p-i-n* structure and the charge separation begins immediately after the photo-excitation, giving rise to the vertical current. To reveal the intrinsic speed limits in this device structure, we consider a delta function-like temporal profile of the excitation pulse for the purpose of our theoretical analysis. The expressions below are derived, based upon the consideration of the expression for the drift current density $j = -en\mu E$ where *e* is the (positive) electron charge, *n* is the photo-excited electron density, $\mu$ is the electron mobility in the *i*-GaAs layer and *E* is the (built-in plus external) electric field intensity. This assumption is not always fulfilled due to ultrafast dynamics of our system whose characteristic time constant is shorter than the electron scattering time. Therefore, the results calculated by this simplified approach will be used solely as an estimation of the order of magnitude of the variables of interest. The vertical current results in a local charge accumulation in the electrodes and a subsequent screening of the electric field which slows down the motion of charges. Taking into account that the initial electron motion occurs in a dense electron-hole plasma, where the electron scattering time decreases approximatively by a factor of 5 [42], we can define a reduced mobility $\bar{\mu}$ ($\bar{\mu} \approx \mu/5 \sim 4000$ cm$^2$/Vs) thus resulting in $j = -e\bar{\mu}nE$. The temporal change of the surface electric charge density $\sigma$ follows the continuity equation and the temporal change of the electric field in the close vicinity of the electrode is then $\dot{E} = \dot{\sigma}/\varepsilon = j/\varepsilon = -e\bar{\mu}nE/\varepsilon$, where $\varepsilon$ is the material electric permittivity. Finally, we can make an estimate of the vertical current relaxation time in the form:

$$T_{\text{vert}} = \frac{\varepsilon}{e\bar{\mu}n}, \quad (S7)$$

assuming $n$ = const. This is not, however, fully true since we did not take into account the full spatio-temporal evolution of both the electron and hole density – the electron density in Eq. (S7) is over-estimated and, therefore, the time constant represents the lower limit. Putting the parameters considered in numerical calculations presented in Supplementary Note 6 to Eq. (S7), we arrive at the estimate of the order of magnitude $T_{\text{vert}} \sim 10$ fs. As this value is smaller than the electron scattering time of 150 fs [42], we conclude that the vertical current dynamics is more complex than a mono-exponential decay assumed above. Nevertheless, the deduced time constant represents still a good estimate of the initial decay order of magnitude for the vertical current pulse. For ultrafast applications, interesting is not only the current but also the total charge transferred during this fast initial current phase. This can be easily estimated based on the charge surface density required to screen the field. For the external bias $U_0 = -10$ V, device height $h = 1$ μm and radius of excited area $R_{\text{exc}} = 2.2$ μm, the number of electrons is $\pi R_{\text{exc}}^2 \varepsilon |U_0|/(eh) \approx 10^5$, which corresponds to charge of $\approx 10^{-14}$ C and current amplitude of up to $\approx 1$ A for the delta function-like laser pulse or $\approx 100$ mA for the 100 fs long laser pulse used in the experiment. Clearly, the very fast initial current impulse has a drawback in a rather small amount of transferred charge and subsequent slow dynamics where the majority of electrons reach the electrode. Note that we analyze only the electrons here: holes have much smaller mobility and, therefore, the initial phase of current is covered solely by the transport of electrons.

It follows from this analysis that the amount of the charge which is transferred within this very initial phase can be controlled by the external voltage and the laser spot size. From the point of view of possible applications, the question is whether we can modify the experimental conditions in order to achieve an ultrafast Oersted magnetic field of a desired peak value and time-integrated magnitude. We can estimate the temporal profile of the initial current in the form:

$$I(t) = \frac{\pi R_{\text{exc}}^2 e \bar{\mu} n |U_0|}{h} \exp[-t/T_{\text{vert}}] \,. \tag{S8}$$

We remind here that the above formula is an estimate of the very initial current peak with ultrafast (subpicosecond) relaxation time that is followed by the further slower system evolution [characterized by a flat part starting at $\sim 10$ ps in Fig. S.6(a)], which will be discussed below and which is not covered by Eq. (S8). It is, nevertheless, possible to optically excite such low particle density that there are no electrons left in the initially excited area after the complete screening buildup. The vertical current would immediately drop down to zero after this ultrafast initial phase which is a scenario perfectly suitable for the ultrafast manipulation of magnetic moments. For example, the Oersted field would then last less than one picosecond only. The external bias $U_0$ is exactly screened out after collection of the photo-created electron density $n_{max}$ into the electrode, where

$$n_{\max} = \frac{\varepsilon |U_0|}{eh^2}. \tag{S9}$$

Inserting this into Eq. (S8) we can express the magnetic field intensity $H$ in the distance $r$ from a current filament as $H=I/2\pi r$. The maximum field amplitude is achieved in a close vicinity of the excited spot, i.e., for $r = R_{\text{exc}}$. Considering the infinitely short delta function-like excitation pulse and performing some algebra, we get the estimates for the peak and integrated magnetic field intensity as:

$$H_{\text{peak}} = \frac{R_{\text{exc}}\bar{\mu}\varepsilon U_0^2}{2h^3} \, , \tag{S10}$$

$$H_{\text{integrated}} = \int_0^\infty H(t)dt = \frac{R_{\text{exc}}|U_0|\varepsilon}{2h} \, . \tag{S11}$$

In a real experiment, however, the time relaxation constant introduced in Eq. (S7) is much shorter than the actual laser pulse duration and the vertical current pulse duration is therefore determined by the temporal profile of the laser pulse. Taking into account parameters of our simulations, external bias $U_0 = -10$ V and optical pulse length 0.1 ps, we estimate (by considering the convolution) that the peak Oersted field is 20 Oe. As Eq. (S10) predicts the peak field of 60 Oe, we may conclude that Eq. (S10) gives a good estimate of the field upper limit for the subpicosecond optical excitation.

The above expressions may be interpreted in a way that we can independently tune the peak Oersted field and its action by properly setting the radius of the excited area and by setting the external bias. Further possibility is to control the peak field by a proper tuning of the excitation density. If excitation density lower than $n_{max}$, which is given by Eq. (S9), is used, a lower peak Oersted field is obtained without a change of the ultrafast field temporal profile. On the other hand, if higher excitation density is used, the peak field would be enhanced as compared to Eq. (S10) but, simultaneously, the magnetic pulse would be longer due to a slow discharging of the structure. Moreover, if very short magnetic pulses are required for a certain application, it is essential to use excitation femtosecond laser pulses as short as possible because their duration would have the dominant effect for the resulting magnetic pulse duration.

*(b) Discharging phase of vertical current*

The slow discharging of the photo-excited *i*-GaAs layer has a complex non-exponential dynamics determined by a spatial distribution of the electric field and charges. The current magnitude is proportional to the external bias but not to the photo-created charge density. Therefore, the temporal width of the vertical current should be proportional to the total amount of excited charge carriers and inversely proportional to the applied bias, as confirmed by our numerical calculations.

As already mentioned above, we can avoid this intermediate slow part of the dynamics by a photo-creation of only a limited density of charged particles, which is described by Eq. (S9). On the other hand, for certain applications, the sub-picosecond duration of the Oersted field pulse might not be necessary and achieving a strong field pulse, with a time integral larger than the value given by Eq. (S11) for a maximum allowed external bias $U_0$, might be advantageous. In such a case, the electron density exceeding $n_{max}$ can be used. Consequently, it results in a slow decay of both the electron density in the *i*-GaAs layer and the vertical current for times exceeding ~ 10 ps, as depicted in Fig. S.7(b) and Fig. S.6(a), respectively. Here, the temporal length of the Oersted field pulse reveals a nontrivial dependency on the external bias. The time-integral of the magnetic field intensity can be expressed from the time integral of the vertical current: as this is the total electron charge in the structure, we may write directly (assuming a negligible electron-hole recombination):

$$H_{\text{integrated}} = \frac{R_{\text{exc}} e n h}{2}. \tag{S12}$$

For the photo-created charge density $3 \times 10^{17}$ cm$^{-3}$ used in our simulations and the external bias $U_0 = -10$ V, the overall time-integrated field is equivalent to the mean field intensity 3.7 Oe acting for 60 ps.

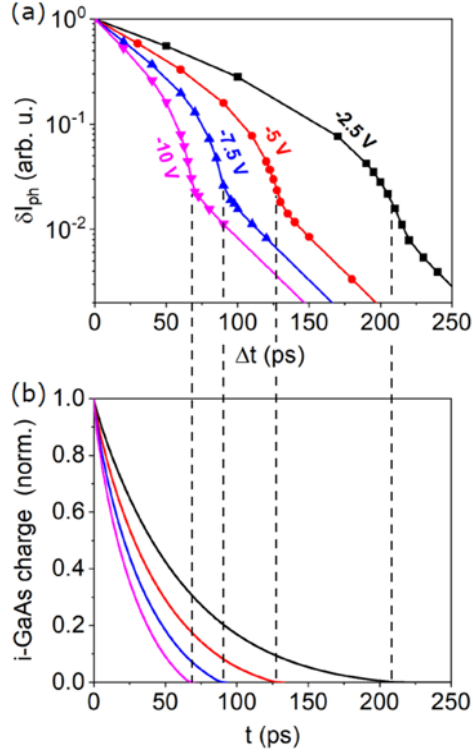

**Fig. S.7**: Demonstration of a dominant role of the *i*-GaAs layer discharging on the photocurrent correlation measurement in the Fe/GaAs *p-i-n* diode. (a) Data from Fig. S.6(d) re-plotted in logarithmic scale. (b) Charge depletion from *i*-GaAs for several applied biases.

*(c) Removal of screening charge by lateral current*

The dynamics at longer time delays, when all the charges are drained out of the photo-active region, is driven by the electric field inside the electrodes acting in the radial direction of the device. Keeping in mind that the width of the device is much larger than its height, we derived an approximate expression for the time constant for the residual charge transport from the device:

$$T_{\text{res}} = \frac{\varepsilon R_0^2}{4eh} \left[ \frac{1}{n_1 t_1 \mu_1} + \frac{1}{n_2 t_2 \mu_2} \right], \tag{S13}$$

where $R_0$ is the device radius, and the parameters of the respective doped layers ($j = 1, 2$) are the acceptor/donor concentration $n_j$, the layer thickness $t_j$ and the majority carrier mobility $\mu_j$. For a multilayered electrode, we sum up contributions to the overall conductivity from all participating layers in the particular denominator in Eq. (S13). This expression illustrates the major contribution of the slower holes to the overall charge dynamics. Consequently, increasing the conductivity of the electron channel does not speed up the overall charge transport. Moreover, this time constant is independent of the external bias, as confirmed by our numerical calculations [see the final stages of the photocurrent correlation curves shown in Fig. S.7(a)].

The time constant $T_{res}$ affects also the current dynamics in the intermediate phase because it determines the rate at which the screening charge is removed. Therefore, fabrication of devices with a smaller diameter and/or higher hole concentration in the *p*-doped layer would lead to the current speedup. Smaller diameter of devices would also help to deliver shorter pulses to the regions close to ring drain contacts, as depicted in Fig. S.6(c). The other promising way to speed-up the current in the device is to increase the *i*-GaAs layer thickness. The time constant $T_{res}$ would decrease and the total charge stored in the device, and eventually released to the electric circuit, would increase due to the increased volume of the photo-active region.

### *(d) Photocurrent correlation measurements*

The principle of a photocurrent correlation measurement is based upon the influence of the excess charge, which is present in the device due to absorption of the first laser pulse, on the properties and propagation of the photocurrent pulse induced by the second laser pulse. First of all, the charge occupying the *i*-GaAs absorbing layer inhibits an additional excitation of electron-hole pairs and, therefore, it reduces the total charge which is injected in an electric circuit. In addition, charges present in the electrodes after their drainage from the optically active *i*-GaAs region influence the correlation curves since they reduce the electric field in the device, slow down the spatial separation of the carriers injected by the second pulse and, finally, they cause an excessive recombination as compared to the situation without any preceding laser pulse. In principle, the correlation curves should reflect all changes in the charge distribution in the device throughout the whole time interval between the photo-creation of charges till its full discharge to an external electric circuit. Therefore, all three distinct regimes of the charge dynamics, which were described above, should be visible in the photocurrent correlation curves. These particular regimes have, however, rather different imprints in the curves and, therefore, they will not be apparent with the same magnitude.

The fast initial buildup of the screening charge is connected with the drainage of carriers from the *i*-GaAs layer. Considering delta function-like optical excitation pulse (i.e., an instantaneous creation of the electron and hole populations), a significant fraction of electrons leaves the *i*-GaAs part of the sample on sub-picosecond time scales [see inset in Fig. S.6(a)]. On the contrary, the holes have much lower mobility and, therefore, they cannot quit the photo-excited area at the time scales given by Eq. (S7). As a result, optical absorption remains saturated even at the end of this initial ultrafast part of the current dynamics and the initial current dynamics is not apparent in the photocurrent correlation curves, as seen both in the experimental [Fig. S.5] and theoretical [Fig. S.6(d)] data. The second part of the dynamics is connected with a simultaneous drainage of electrons and holes from the optically active *i*-GaAs layer in the device structure. Each drained electron-hole pair allows absorption of an additional photon from the second laser pulse. Due to this, this process is visible as an increase of the total photocurrent in the photocurrent correlation curves. In fact, this part of the dynamics, when the *i*-GaAs is being discharged [see Fig. S.7(b)], covers the main part of the correlation curve [see Fig. S.7(a)]. In the last stage of the current dynamics, when charges have been drained from *i*-GaAs and when they migrate in the radial direction, the photocurrent correlation curves display a joint effect of the screening by the residual charges in the electrodes and of electron-hole recombination. As the amount of the screening charge is only a few percent of the total photo-excited charge in our simulations, the last part of the correlation has only a very little relative magnitude [see Fig. S.7(a)] and could be hidden by a noise in the experimental data.